%% file: paper.tex
\documentclass{sig-alternate-ipsn12}

\usepackage{acronym}
\usepackage{algorithm}
\usepackage{algorithmic}
\usepackage{amsfonts}
\usepackage{amsmath}
\usepackage{amssymb}
\usepackage{color}
\usepackage{epsfig}
\usepackage{float}
\usepackage{graphicx}
\usepackage{ifthen}
\usepackage[latin1]{inputenc}
\usepackage{listings}
\usepackage{multirow}
\usepackage{psfrag}
\usepackage{slashbox}
\usepackage{subfigure}
\usepackage{url}
\usepackage{times}

\title{Jam-X: Wireless Agreement under Interference}

\numberofauthors{1}
\author{
	\alignauthor 
	Carlo Alberto Boano$^\dagger$, Kay R\"{o}mer$^\dagger$, Marco Antonio Z{\'u}ñiga$^\S$, and Thiemo Voigt$^\ddagger$
	\vspace{1.5mm}
\and
	\affaddr{$^\dagger$Institute of Computer Engineering, University of L\"{u}beck, Germany}\\
\and
	\affaddr{$^\S$Networked Embedded Systems Group, University of Duisburg-Essen, Germany}\\
\and
	\affaddr{$^\ddagger$Swedish Institute of Computer Science, Kista, Sweden}\\	
}

\toappear{Copyright is held by the author/owner(s).\\Corresponding author: cboano@iti.uni-luebeck.de\\}

\begin{document}
\maketitle

\input{abstract}




\input{introduction}

\input{problem}
\input{jamming}

\input{jamx_unicast}

\input{evaluations_unicast_2}

\input{evaluations_unicast_3}
\input{jamx_broadcast}
\input{related}

\input{conclusions}

\fontsize{8pt}{1ex}\selectfont
\bibliographystyle{unsrt}
\bibliography{ref}

\end{document}

%% file: abstract.tex
\begin{abstract}

Wireless low-power transceivers used in sensor networks such as IEEE~802.15.4 typically
operate in unlicensed frequency bands that are subject to external interference from 
devices transmitting at much higher power. Communication protocols should
therefore be designed to be robust against such interference.
A critical building block of many protocols at all layers is \emph{agreement} on a 
piece of information among a set of nodes. At the MAC layer, nodes may need to agree
on a new time slot or frequency channel; at the application layer nodes may need
to agree on handing over a leader role from one node to another. Message loss
caused by interference may break agreement in two different ways: none of the
nodes use the new information (time slot, channel, leader) and stick with the
previous assignment, or -- even worse -- some nodes use the new information and
some do not. This may lead to reduced performance or failures.

In this paper we investigate the problem of agreement under interference and
point out the limitations of the traditional message-based n-way handshake
approach. We propose novel protocols that use jamming instead of message
transmissions and show that they outperform the n-way handshake 
in terms of agreement probability, energy consumption, and time-to-completion
both in the unicast case (two neighboring nodes agree) as well as in the
broadcast case (any number of neighboring nodes agree).


\end{abstract}

%% file: introduction.tex
\section{Introduction} \label{sec:introduction}

Wireless sensor nodes often need to agree on fundamental pieces of information
that can drastically affect the performance of the network. Several
state-of-the-art MAC protocols use time division multiple access (TDMA) or
frequency diversity techniques to optimize their performance, in order to
maximize network lifetime and minimize battery depletion. In such protocols,
vital information such as the TDMA schedule, the channel-hopping sequence
derived by interference-aware protocols, or the seed used to regulate the
random channel hopping need to be agreed upon by two or more sensor nodes in a reliable
fashion. Failure to agree on such information correctly (e.g., nodes using
inconsistent TDMA schedules) may lead to a disruption of the network
connectivity or to a substantial performance degradation.

When sharing information using an unreliable medium (such as wireless), no
delivery guarantee can be given on the messages that are sent. Akkoyunlu et
al.~\cite{akkoyunlu75constraints} have shown that, in an arbitrary distributed
facility, it is impossible to provide the so called \emph{complete status},
i.e., one cannot guarantee that two distributed parties know the ultimate fate
of a transaction and whether they are in agreement with each other.

The problem is further exacerbated in the presence of external interference: the
low-power transmissions of wireless sensor networks are highly vulnerable to
interference caused by radio signals generated by devices operating in the
same frequency range. Several studies have highlighted the increasing congestion
of the unregulated ISM bands used by sensornets to communicate, especially the
2.4 GHz band~\cite{zhou06crowded}.
Sensornets operating on such frequencies must cope with simultaneous
communications of WLAN and Bluetooth devices, as well as with the
electromagnetic noise generated by domestic appliances such as microwave ovens.

Hence, it is important to derive reliable techniques to ensure
agreement among nodes and make sure that they are robust to external
interference. At the same time, these techniques need to be efficient
since sensor nodes have limited computational capabilities and energy
resources.


Traditional communication protocols make use of one or more acknowledgment
messages to verify whether the information was successfully shared (e.g., TCP
handshake). We show that this approach is not optimal under external
interference for two main reasons. Firstly, the probability of receiving a
packet correctly is small, and therefore the chances of receiving successfully
an entire sequence of acknowledgment packets is even lower. Secondly, the
overhead introduced by the packet header and footer is much larger than the
information carried by the acknowledgment message itself, making it
unnecessarily more vulnerable to interference.

In this work, we design, implement, and evaluate Jam-X, a family of simple yet
efficient agreement protocols for wireless sensor networks challenged by
external interference. Instead of exchanging acknowledgment packets, Jam-X
introduces a jamming sequence to inform the other node(s) about the correct
reception of a message carrying the information to be agreed upon.
The key insight behind this approach is that jamming can be reliably detected in
the presence of interference, while packets are destroyed by interference.

Jam-X is intended as a building block to construct protocols at different layers
of the protocol stack. It could be embedded into a MAC protocol to agree on time
slots or frequency channels, at the transport level to agree on connection
establishment or teardown, or at the application level to agree on handover of
a leader role from one node to another.

We design Jam-X variants for both unicast (two neighboring nodes need to agree)
and broadcast (an arbitrary number of neighboring nodes need to agree)
scenarios, and show that they outperform traditional packet-based handshake
protocols in the presence of heavy interference with respect to agreement
probability, energy consumption, and time-to-completion. In particular, the time
to complete the agreement is
predictable, making Jam-X especially suited for applications with timeliness
requirements, for example, when using Jam-X as a building block to construct a
MAC protocol.



Our paper proceeds as follows. Section~\ref{sec_problem} defines the agreement
problem in wireless networks with particular emphasis on external interference.
Section~\ref{sec:jamming} conveys the main idea of the paper: using jamming as a
binary signal for acknowledging packets. Section~\ref{sec:jamx_unicast}
introduces Jam-X, a protocol for unicast agreement. In
Section~\ref{sec:evaluation_unicast} we present our experimental results and
compare the performance of Jam-X and packet-based handshakes.
Section~\ref{sec:jamx_broadcast} introduces Jam-B, a protocol for broadcast
agreement under interference, and analyzes its performance experimentally. We
review related work in Section~\ref{sec:related} and conclude our paper in
Section~\ref{sec:conclusions}.

%% file: problem.tex
\section{Problem Definition} \label{sec_problem}

To agree on a given piece of information is a classical coordination problem in
distributed computing. The \emph{Two Generals' Agreement Problem}, formulated
for the first time by Jim Gray~\cite{gray78database} to illustrate the two-phase
commit protocol in distributed database systems, is often used to explain the
design challenges when attempting to coordinate an action by communicating over
a faulty channel, and can be described as follows.

Two battalions are encamped near a city, ready to launch the final attack.
Because of the redoubtable fortifications, the attack must be carried out by
both battallions at the same time in order to succeed. Hence, the generals of
the two armies need to agree on the time of the attack, and their only way to
communicate is to send messengers through the valley. The latter is occupied by
the city's defenders, and a messenger can be captured and its message lost, i.e.,
the communication channel is unreliable. Since each general must be aware that
the other general has agreed on the attack plan, messengers are used also to
exchange acknowledgments. However, because the acknowledgement of a message
receipt can be lost as easily as the original message, a potentially infinite
series of messages is required to reach an agreement. A different problem that
we are not addressing in this work is how to guarantee the identity of the
sender of the message, as well as how to cope with misbehaving parties.

\subsection{Agreement in Wireless Networks} \label{subsec:agreement_wireless}

\begin{figure}[t]
	\begin{center}
		\includegraphics[width=0.46\textwidth]{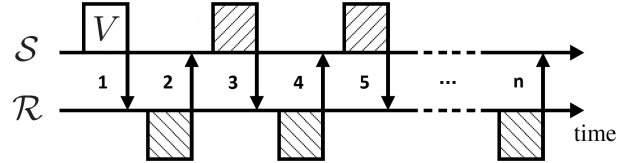}
		\vspace{-2mm}
		\caption{N-way handshake between nodes $\mathcal{S}$ and $\mathcal{R}$.}
		\vspace{-6mm}
 		\label{fig:drawing_handshake}
	\end{center}
\end{figure}

\begin{figure*}[t]
	\begin{center}
		\subfigure[Positive Agreement]{
			\includegraphics[width=0.32\textwidth]{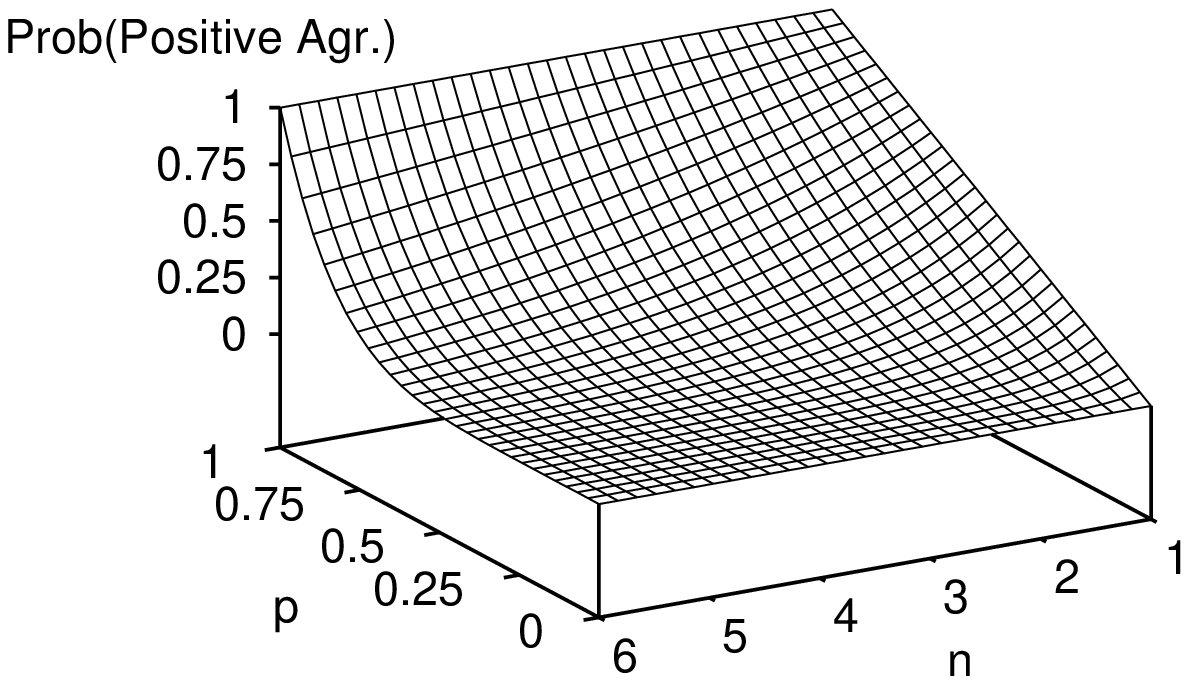}
			\label{fig:p_positive_agreement_3d}
		}
		\subfigure[Negative Agreement]{
			\includegraphics[width=0.32\textwidth]{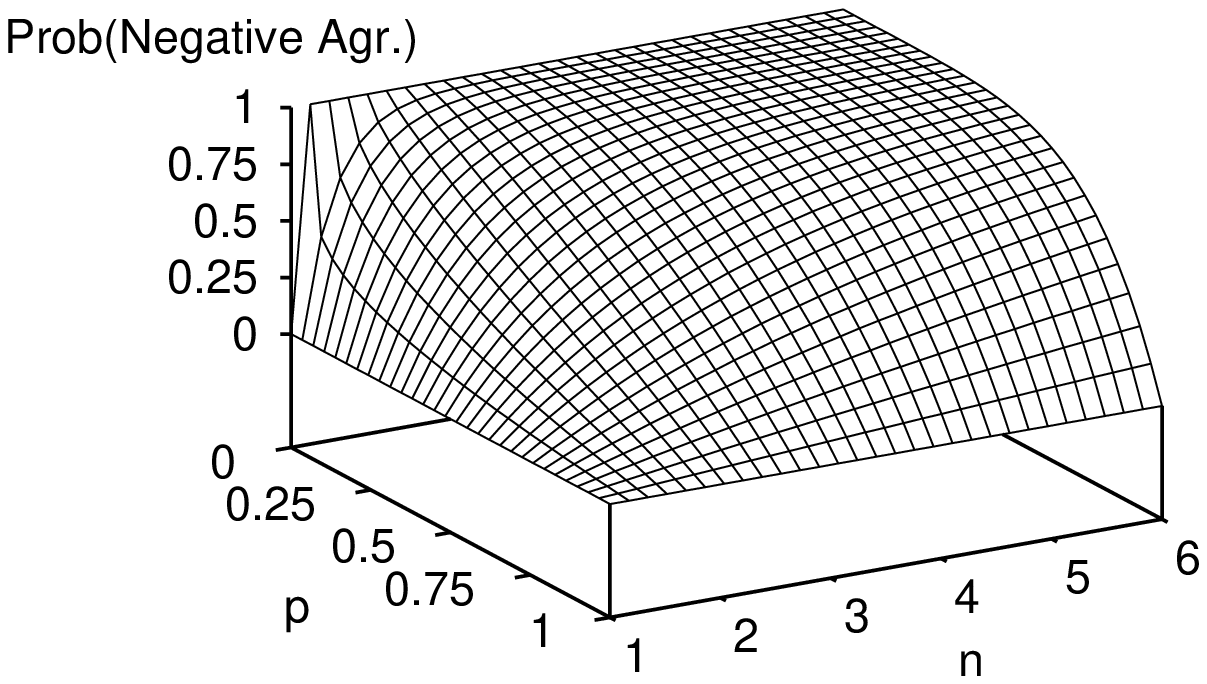}
			\label{fig:p_negative_agreement_3d}		
		} 		
		\subfigure[Disagreement]{
			\includegraphics[width=0.32\textwidth]{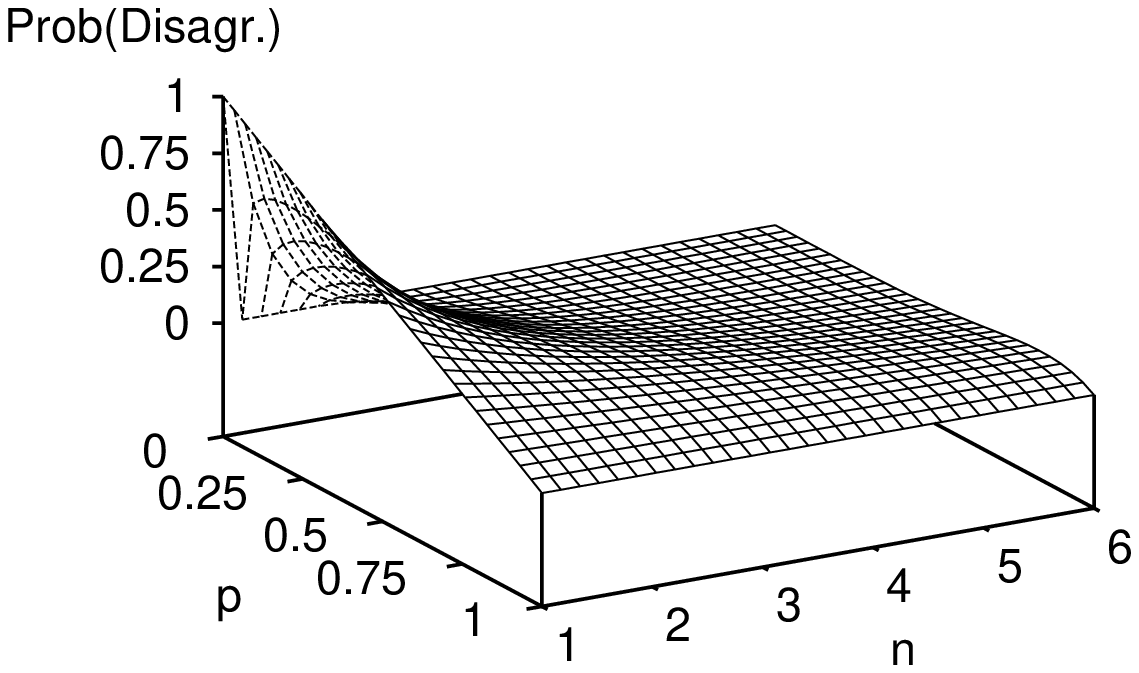}
			\label{fig:p_disagreement_3d}		
		}
		\vspace{-4mm}
		\caption{Distribution of the probabilities of positive/negative agreement and
		disagreement, as a function of the probability of successful transmission $p$
		and amount of exchanged packets $n$. Note the different orientation of the axes.}
		\vspace{-5mm}
 		\label{fig:probabilities_3d}
	\end{center}
\end{figure*}

In the context of wireless communications, the problem can be rephrased as
follows. When two nodes, $\mathcal{S}$ and $\mathcal{R}$, need to agree on a
common value $V$, they exchange a sequence of $n$ messages in an alternating
manner (Figure~\ref{fig:drawing_handshake}). Node $\mathcal{S}$ is the initiator
of the exchange. After the transmission of $V$, each message acknowledges the
receipt of the former message, i.e., a node sends message $i>1$ only if it
correctly received message $i\!-\!1$. There are no retransmissions, i.e., if a
message is lost, the exchange is terminated. Each node uses a simple rule to
determine the success of the exchange: if all expected messages are received,
the exchange is deemed successful, otherwise the exchange is deemed
unsuccessful.

Please note that this scenario corresponds to an \emph{n-way handshake} between
nodes $\mathcal{S}$ and $\mathcal{R}$, where $n$ is the number of packets
exchanged. The n-way handshake is a widely used mechanism in communication
networks. For example, TCP employs a 3-way handshake ($n\!=\!3$) to establish
connections over the network, whereas IEEE 802.11i (WPA2) uses a 4-way handshake
($n\!=\!4$) to carry out the key exchange.

This scenario leads to three possible outcomes:
\vspace{-2mm}
\begin{enumerate}
  \item \textbf{Positive Agreement.} The $n$ messages are all received
  correctly, and both nodes deem the exchange as successful, accepting $V$.
  \vspace{-2mm}
  \item \textbf{Negative Agreement.} A message $m$ with $m<n$, i.e., a message
  prior to the last message $n$, is lost. None of the two nodes receives all the
  expected messages, hence both nodes deem the exchange as unsuccessful,
  discarding $V$.
  \vspace{-2mm}
  \item \textbf{Disagreement.} The message $n$, i.e., the last message, is lost.
  One of the two nodes receives all the expected messages, and deems the
  exchange as successful, while the second node misses the last message and
  therefore deems the exchange as unsuccessful.
\end{enumerate}
\vspace{-0.5mm}

In the original two generals' scenario, a \emph{positive agreement} would lead
to a simultaneous attack of the city by both battalions and a consequent
victory, a \emph{negative agreement} would cause both battalions to stall, while
a \emph{disagreement} would trigger the attack of only one battalion, and a
consequent defeat of the attacking forces.

Notice that while a \emph{disagreement} is a potentially pernicious outcome, a
\emph{negative agreement} is often less severe. For example, if the shared value
contains the next wireless channel to be used for communication, two nodes are
better staying in the same lossy wireless channel, rather than having only one
of them move to a different channel. The probability of negative agreements
should, however, be minimized, as it may lead to reduced performance.



The frequency of the three outcomes described above is strongly dependent on the
link quality. Letting $p$ represent the probability that a generic message is
successfully received (assuming that $p$ remains constant over time and that it
is independent for each packet), and $n$ the length of the n-way handshake, we
obtain:
\vspace{-0.5mm}
\begin{eqnarray}
\text{Prob} ( Positive Agreement ) &=& p^n \nonumber \\
\text{Prob} ( Negative Agreement ) &=& 1 - p^{n-1} \nonumber \\
\text{Prob} ( Disagreement ) &=& p^{n-1} (1-p) \nonumber
\label{eq.prob}
\end{eqnarray}
\vspace{-4mm}

Figure~\ref{fig:probabilities_3d} illustrates the distribution of the
probabilities of positive agreement (a), negative agreement (b), and
disagreement (c), as a function of the probability $p$ of successful packet
transmission and length of the n-way handshake $n$. The plots offer two
important insights. Firstly, increasing the communication quality leads to more
agreements: when $p$ is high, the probability of positive agreement is maximized.
Secondly, long n-way handshakes minimize the chances of disagreement: when $n$
is high, the probability of disagreement is minimal.

Besides minimizing the chances of disagreement, long n-way handshakes also
increase the probability of negative agreement, especially when dealing with
unreliable channels. For this reason, minimizing the amount of disagreements
while maximizing the amount of positive agreements becomes a catch-22 dilemma in
the presence of unreliable links. Figure~\ref{fig:probabilities_3d} shows that long
n-way handshakes minimize \textbf{both} the probability of disagreement and the
probability of positive agreement, while short n-way handshakes maximize
\textbf{both} the probability of positive agreement and the chances of
disagreement.


\begin{figure*}[t]
	\begin{center}
		\subfigure[IEEE~802.15.4 packets]{
			\includegraphics[width=0.235\textwidth]{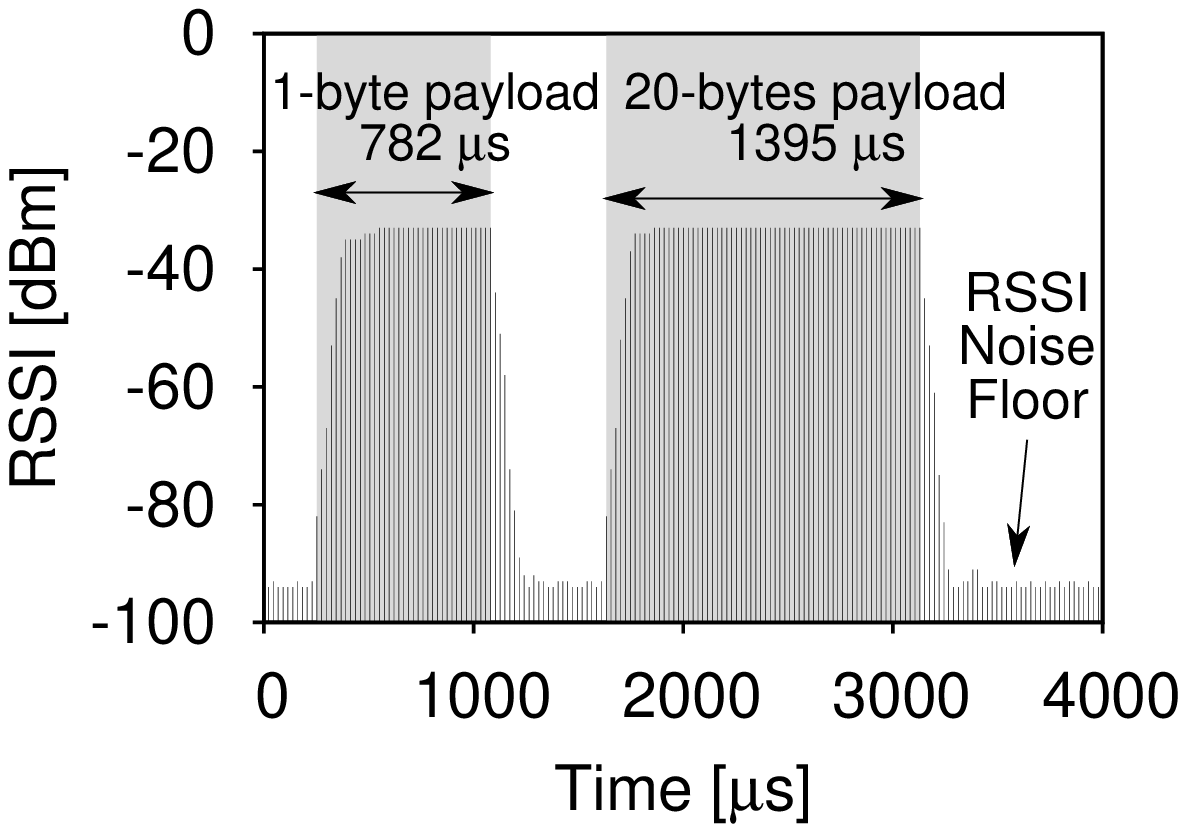}
			\label{fig:zigbee}		
		}	
		\subfigure[Heavy Wi-Fi interf.]{
			\includegraphics[width=0.235\textwidth]{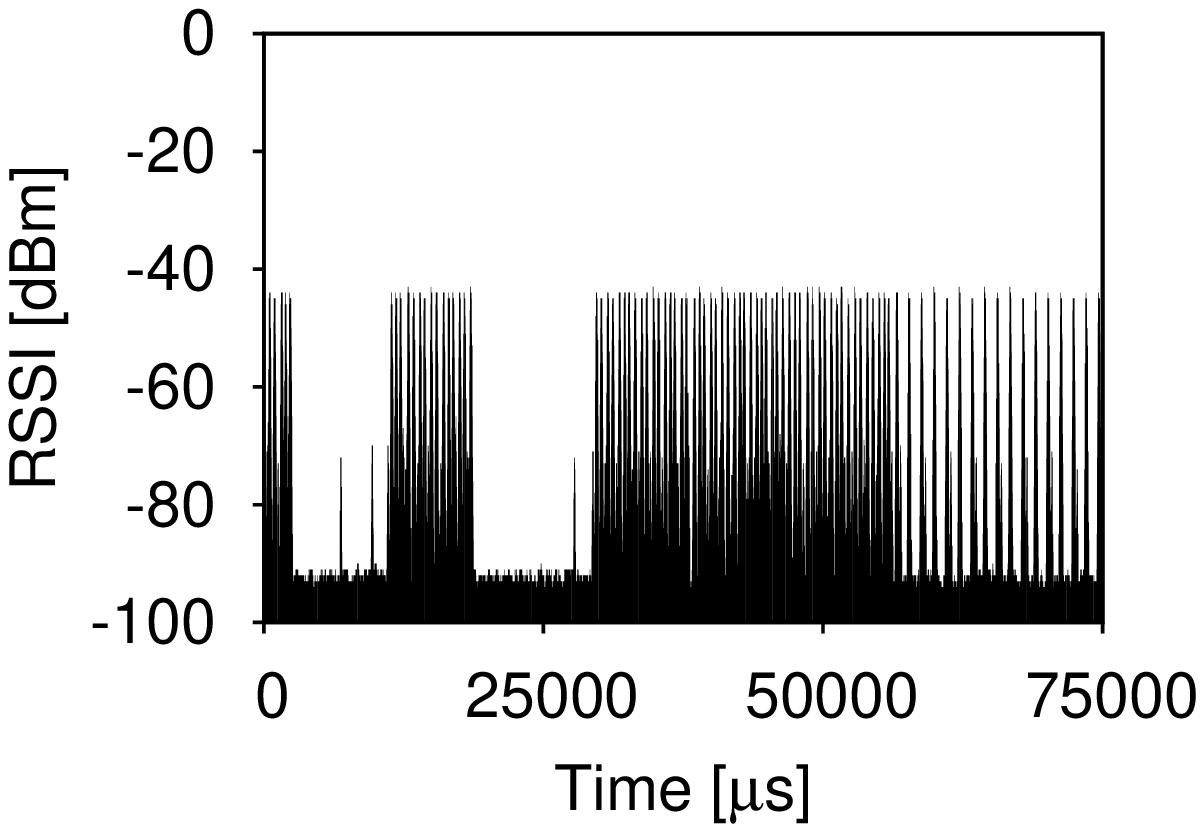}
			\label{fig:heavy_wifi_1}		
		}	
		\subfigure[Bluetooth interf.]{
			\includegraphics[width=0.235\textwidth]{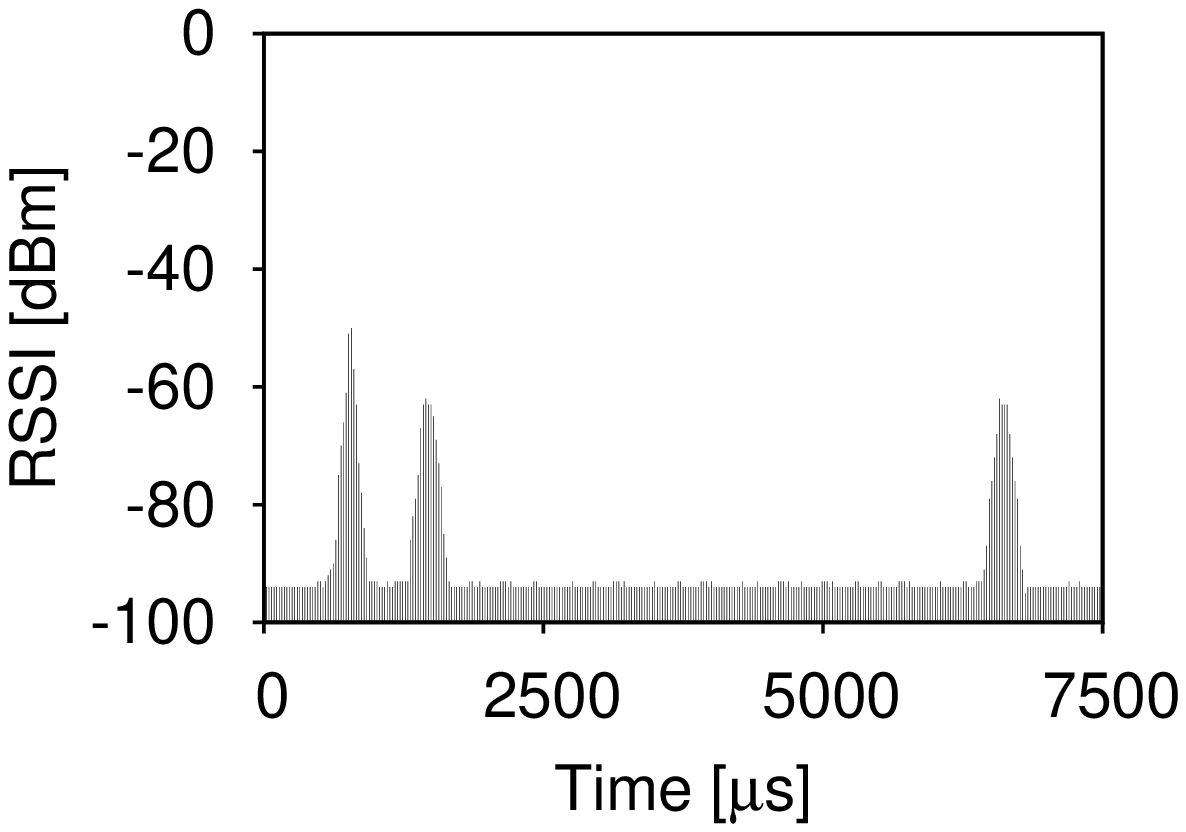}
			\label{fig:bluetooth}
		}		
		\subfigure[Microwave oven interf.]{
			\includegraphics[width=0.235\textwidth]{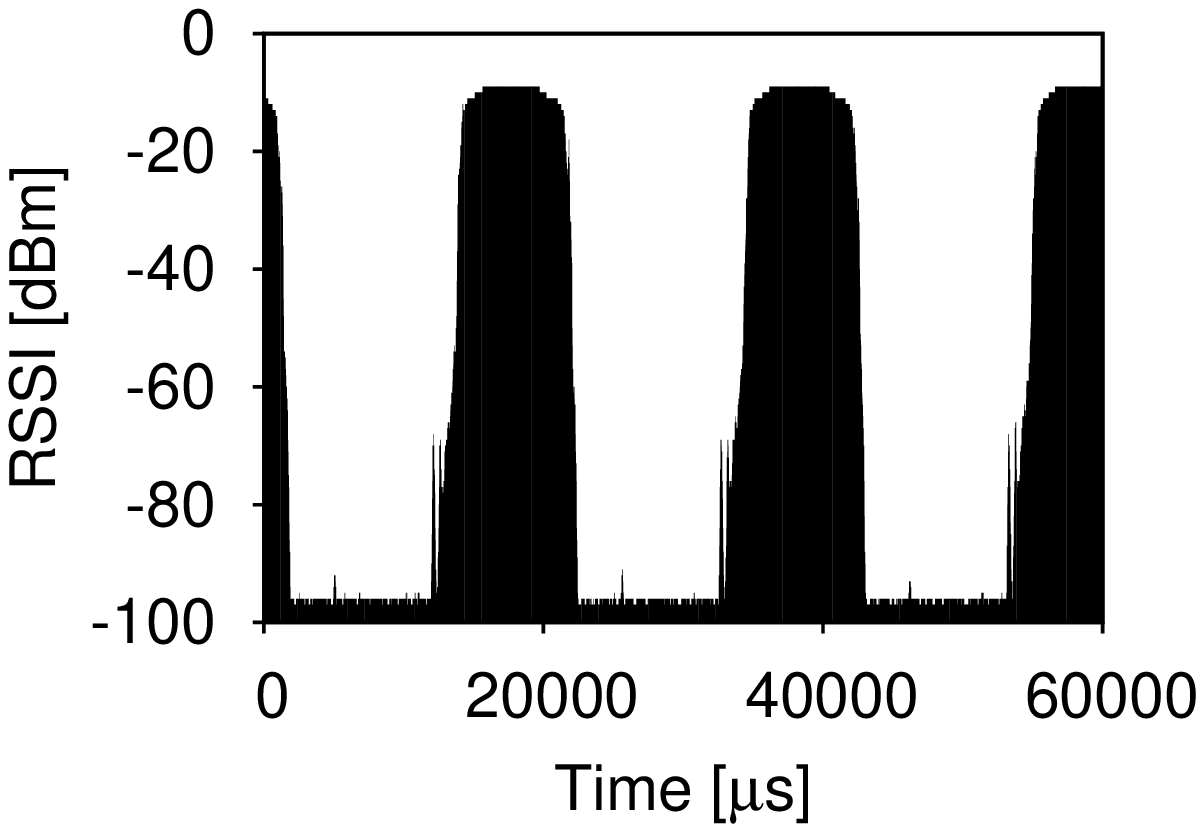}
			\label{fig:oven}		
		}		
		\vspace{-3mm}				
		\caption{RSSI values measured using off-the-shelf sensor motes operating in the 2.4 GHz
		ISM band.}
		\vspace{-6mm}
 		\label{fig:rssi_traces_1}
	\end{center}
\end{figure*}

\vspace{-2mm}
\subsection{Agreement in Wireless Sensor Networks Challenged by External Interference} \label{subsec:agreement_wsn}

In the context of wireless sensor networks, minimizing the amount of exchanged
packets $n$ is mandatory because of the limited energy resources available on
sensor nodes, i.e., one needs to minimize the time in which the radio is active
as much as possible.
Hence, one should try to maximize the probability $p$ of successful packet
exchanges instead. Redundant packet transmissions (i.e., repeating a handshake
message several times and assuming successful transmission if at least one copy
is received) can achieve this, but are not advisable as they also increase
energy consumption.

Another aspect that affects $p$ is the channel quality itself. Unfortunately,
sensor motes operate in unlicensed ISM radio bands, and often use a very low
transmission power, which makes them vulnerable to external interference. Any
wireless appliance operating in the same frequency range of sensornets can
potentially interfere with their communications and therefore radically decrease
the probability of successful packet exchange $p$. In the 2.4 GHz ISM band, for
example, Wi-Fi and Bluetooth networks as well as domestic appliances such as
microwave ovens can create noise levels that overwhelm the interference
resistance capabilities of DSSS radios and radically decrease the packet
reception rate~\cite{zhou06crowded, musaloiu08minimising}.


Hence, we will investigate ways to encode transmissions such that their success
probability $p$ is maximized for interfered channels. However, for this we first
have to understand the charateristics of interfered channels.

\subsection{Analysis of Common Interference Sources} \label{subsec:interference_analysis}

In order to understand and mitigate the impact of external interference on the
probability of successful transmission $p$ in wireless sensor networks
communications, we study the interference patterns produced by common devices
operating in the 2.4 GHz ISM band. We perform a high-speed sampling of the RSSI
register of the CC2420 radio ($\approx 50$~kHz as in~\cite{boano11jamlab})
using Maxfor MTM-CM5000MSP and Sentilla Tmote Sky motes. We call this operation
\emph{fast RSSI sampling} over a time window $t_{samp}$. 
Figure~\ref{fig:rssi_traces_1} shows the outcome of fast RSSI sampling both
in the presence of sensornet communications and external interference. \\

\vspace{-2.5mm}
\textbf{Absence of external interference.} 
When neither interference nor IEEE~802.15.4 packet transmissions are present,
the fast RSSI sampling returns the so called noise floor. The latter has
typically values in the proximity of the radio sensitivity threshold (e.g., in
the range $[-100,-94]$ dBm for the CC2420). In the presence of IEEE~802.15.4
packet transmissions, fast RSSI sampling returns values that correspond to the
received signal strength upon packet reception (the same value that would be
returned by the RSSI indicator) lasting for the duration of the transmission
(Figure~\ref{fig:zigbee}). As packets have a constrained maximum payload size of
127 bytes according to the 802.15.4~PHY standard, a packet transmission with 250
Kbit/sec will not last more than $\approx$ 4.3 ms. \\

\vspace{-2.5mm}
\textbf{Presence of external interference.} In the presence of interference, fast
RSSI sampling can return different outcomes that depend on several variables.
For example, Wi-Fi traffic depends on the number of active users and their
activities, and on the traffic conditions in the backbone.

Figure~\ref{fig:heavy_wifi_1} shows the outcome of fast RSSI sampling in the
presence of heavy Wi-Fi interference (caused by a file transfer). The
transmissions of Wi-Fi devices are much stronger than 
sensornet transmissions, and can affect several IEEE~802.15.4 channels at the
same time. Hauer et al.~\cite{hauer09interference, hauer10ewsn} have shown that
with a sufficiently high sampling rate, one can identify the short instants in
which the radio medium is idle due to the Inter-Frame Spaces (IFS) between
802.11 packets. Figure~\ref{fig:heavy_wifi_1} shows that it is indeed possible
to identify RSSI values matching the radio sensitivity threshold between
consecutive Wi-Fi transmissions.

Figure~\ref{fig:bluetooth} shows an example of interference generated by
Bluetooth. The latter uses an Adaptive Frequency Hopping mechanism to combat
interference, and hops among 1-MHz channels around 1600 times/sec., hence it
remains in a channel for at most 625~$\mu$s. Since Bluetooth channels are more
narrow than the ones defined by the 802.15.4 standard, it may happen that
communication in multiple adjacent Bluetooth channels affects a single 802.15.4
channel.

Figure~\ref{fig:oven} shows an example of the interference pattern caused by
microwave ovens. Microwave ovens emit high-power noise ($\approx$~60 dBm) in the
2.4 GHz frequency band in a very periodic fashion: the duration of
their interference does not last more than $\approx$~10 ms before being followed
by $\approx$~10 ms quiescence~\cite{boano11jamlab}. \\

\vspace{-2.5mm}
\textbf{The role of idle periods.} In the presence of interference, $n$-way
handshakes need to take advantage of idle periods. In principle, the longer the
idle period and the shorter the exchange sequence, the higher the likelihood of
obtaining positive agreements.
However, the interplay between idle periods and $n$-way handshakes is complex
because of the particular pattern of each interfering source. Some devices, such
as microwave ovens, present periodic patterns with relatively long idle periods,
while others, such as Wi-Fi stations, can present shorter idle periods with a
highly variable and random length. For our purposes, we characterize an
interfering source based on the random variables $IDLE_{i}$ and $BUSY_{i}$,
which denote the distributions of the idle and busy periods.

Our aim is to encode part of the handshake messages such that they can be
successfully received despite interference and for different distributions of
idle and busy periods as detailed in the subsequent section.

%% file: jamming.tex
\section{Jamming as Binary ACK Signal} \label{sec:jamming}

The n-way handshake shown in Figure~\ref{fig:drawing_handshake}
conveys the information $V$ to be agreed upon only in the first
message, whereas the remaining messages are only used to acknowledge
its reception. The acknowledgment packets carry essentially only two
pieces of information: (i) the confirmation of the receipt and (ii)
the identity of sender and receiver.

Based on the discussion in the previous section, all messages should be as short
as possible to increase the chances of fitting into idle periods. With respect
to this, the encoding of acknowledgements in IEEE~802.15.4 messages is very
inefficient. While their payload consists of a single ACK bit, the whole packet
consists of synchronization preamble and a physical header (4-bytes preamble,
1-byte SFD, 1-byte length field) as well as a MAC header (2-bytes frame control,
1-byte sequence number, 4-20-bytes address, 2-bytes FCS, 0-14 bytes auxiliary
security header). If any of the bits in the headers and preamble are corrupted
by external interference, the packet is often
undecodable~\cite{liang10surviving, jamieson07ppr}.

Therefore, instead of encoding acknowledgements as packet transmissions, we
propose to encode acknowledgements by \textbf{jamming}, where the presence of
jamming signals the receipt of the previous message. The key advantage of this
approach is that jamming as generated by off-the-shelf mote radios can be
reliably detected even under heavy interference as we will show in the remainder
of this section.

While a jamming signal can encode the binary acknowledgement information, it
cannot encode the identities of the sender and receiver as a regular packet
would. However, the identities of sender and receiver are included in the first
message carrying the information $V$ to be agreed upon, so there is no need to
repeat this information in the subsequent acknowledgement messages.

All that is needed is that the communication channel remains allocated
exclusively to the sender and receiver of the first message long enough to
conclude the exchange of acknowledgements in form of jamming signals. As during
this period only sender and receiver can access the channel, their identities
are implicitly known to each other. Any protocol that embeds Jam-X as a building
block for agreement needs to meet this requirement. At the MAC layer, RTS/CTS
can be used to allocate the channel in CSMA protocols, in TDMA protocols the
timeslots must be long enough. Note that the number and duration of the jamming
signals is fixed and known in advance as discussed further below.




\begin{figure}[t]
	\begin{center}
		\subfigure[Absence of Interference]{
			\includegraphics[width=0.225\textwidth]{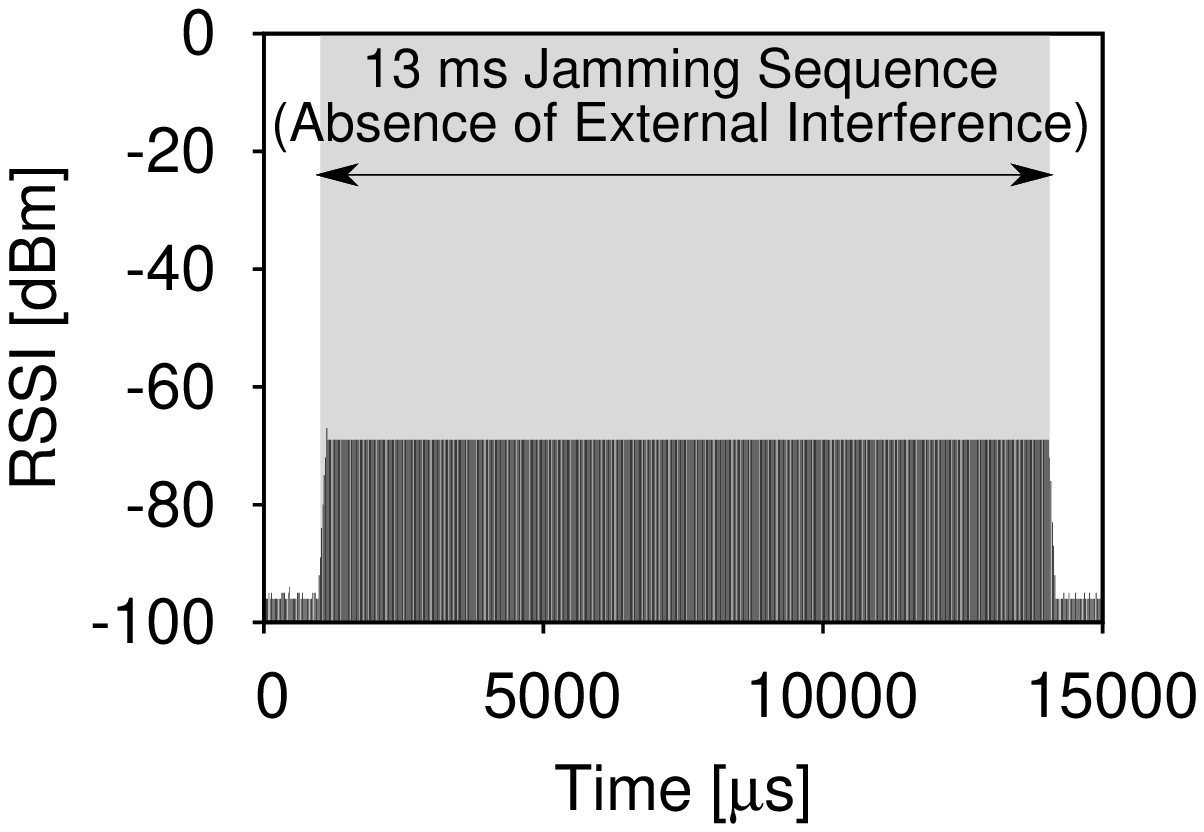}	
			\label{fig:jammer}		
		} 	
		\subfigure[Presence of Interference]{
			\includegraphics[width=0.225\textwidth]{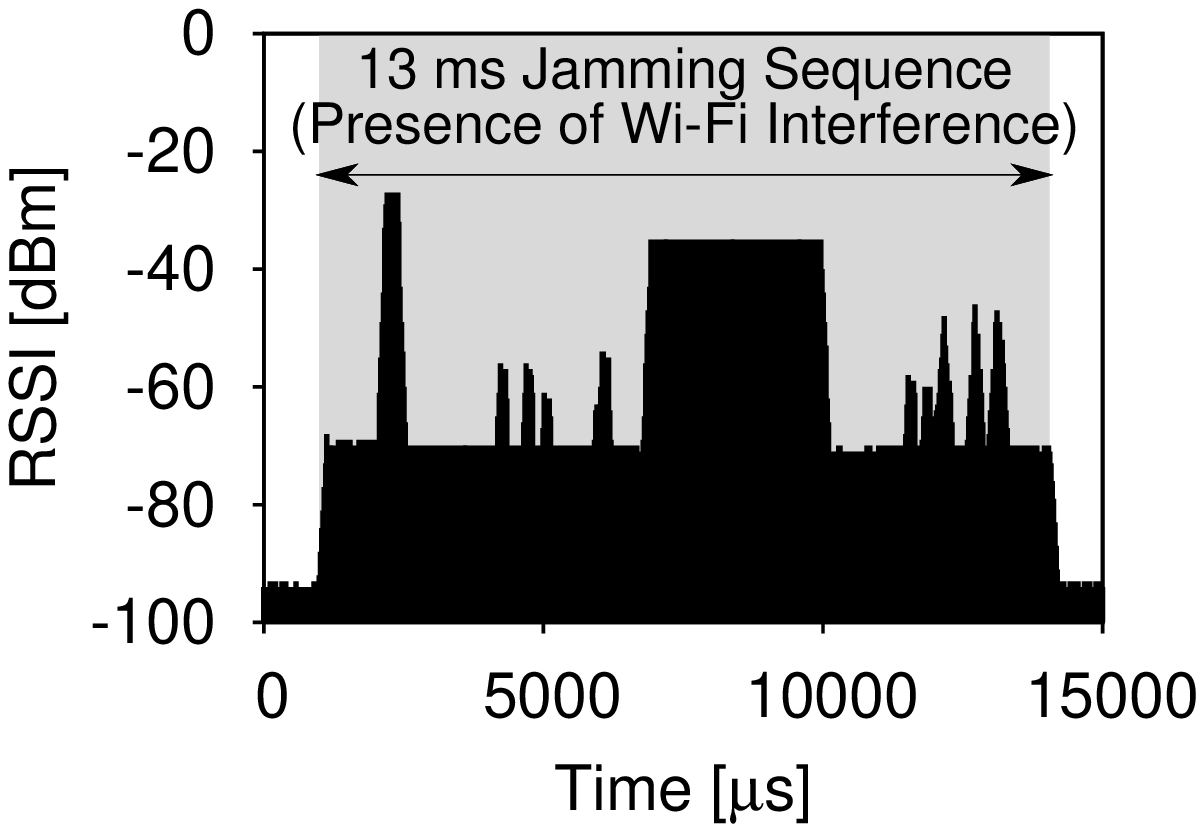}
			\label{fig:jammer_interference}		
		} 		
		\vspace{-3.5mm}
		\caption{RSSI values recorded during the transmission of a jamming
		sequence without external interference (a), and with external Wi-Fi
		interference (b).}
		\vspace{-7mm}
 		\label{fig:scanner_jamming}
	\end{center}
\end{figure}

\vspace{-1mm}
\subsection{Generating a Jamming Sequence} \label{subsec:jamming_generation}
\vspace{-1mm}
Recent work has shown that off-the-shelf radios can be used to
generate controllable and repeatable jamming signals in specific
IEEE~802.15.4 channels by transmitting a modulated or unmodulated
carrier signal that is stable over time~\cite{boano11jamlab,
boano09interference}. This approach is superior to packet-based
jamming, as the generated carrier signal is independent of both packet sizes and
inter-packet times.

We use this approach to generate precisely timed jamming signals.
In particular, we use the popular CC2420 radio
that can generate a continuous modulated carrier signal by properly configuring
the MDMCTRL1 register. The modulated signal can be seen as a pseudorandom
sequence created using the CRC generator, and can be prefixed by a
synchronization preamble such that a receive interrupt is triggered at the
receiving radio. However, we do not use this feature to detect the jamming
sequence. Instead, we sample the RSSI register at high frequency as discussed in
the following section.

\vspace{-1mm}
\subsection{Detecting a Jamming Sequence} \label{subsec:detecting}
\vspace{-1mm}
As already mentioned in Section~\ref{subsec:agreement_wsn}, common radio chips
offer the possibility to read the RSSI in absence of packet transmissions -- the
so called RSSI noise floor measurement or energy detection feature. Several
researchers have shown that it is a useful way to assess the noise and the level
of interference in the environment~\cite{musaloiu08minimising, hauer10ewsn,
polastre04bmac}: RSSI readings close to the sensitivity threshold of the radio
indicate absence (or a limited amount) of interference, while values above this
threshold identify a packet transmission, or a busy/congested medium
(Figure~\ref{fig:rssi_traces_1}).


We sample the RSSI register at high frequency to detect the presence
or absence of a jamming signal generated by a sensor node. As shown in
Figure~\ref{fig:jammer}, a jamming sequence lasting for a time window
$t_{jam}$ results in a stable RSSI value above the sensitivity
threshold of the radio, very similar to a packet transmission (shown in
Figure~\ref{fig:zigbee}). Therefore, if any of the RSSI samples equals
the sensitivity threshold of the radio, no jamming signal is present.

In the presence of additional external interference, the RSSI register
will return the maximum of the jamming signal and the interference
signal due to the co-channel rejection properties of the radio
\cite{boano11jamlab}.  Figure~\ref{fig:jammer_interference}
illustrates this for a jamming signal with simultaneous Wi-Fi
interference. As we have shown in
Section~\ref{subsec:interference_analysis}, typical interference
sources -- in contrast to our jamming signal -- do not produce
continuous interference for long periods of time, rather they
alternate between idle and busy. That is, if the jamming signal lasts
longer than the longest busy period of the interference signal, we are
still able to detect the absence of the jamming signal by checking if
any of the RSSI samples equals the sensitivity threshold of the
radio. 


%% file: jamx_unicast.tex
\section{Jam-X: Wireless Agreement\\ under Interference} \label{sec:jamx_unicast}

In this section, we present Jam-X, a family of jamming-based protocols
designed to provide a high probability of agreement in the presence of
heavy external interference.

\begin{figure*}[t!]
	\begin{center}
		\subfigure[Ack-2]{
			\includegraphics[width=0.2675\textwidth]{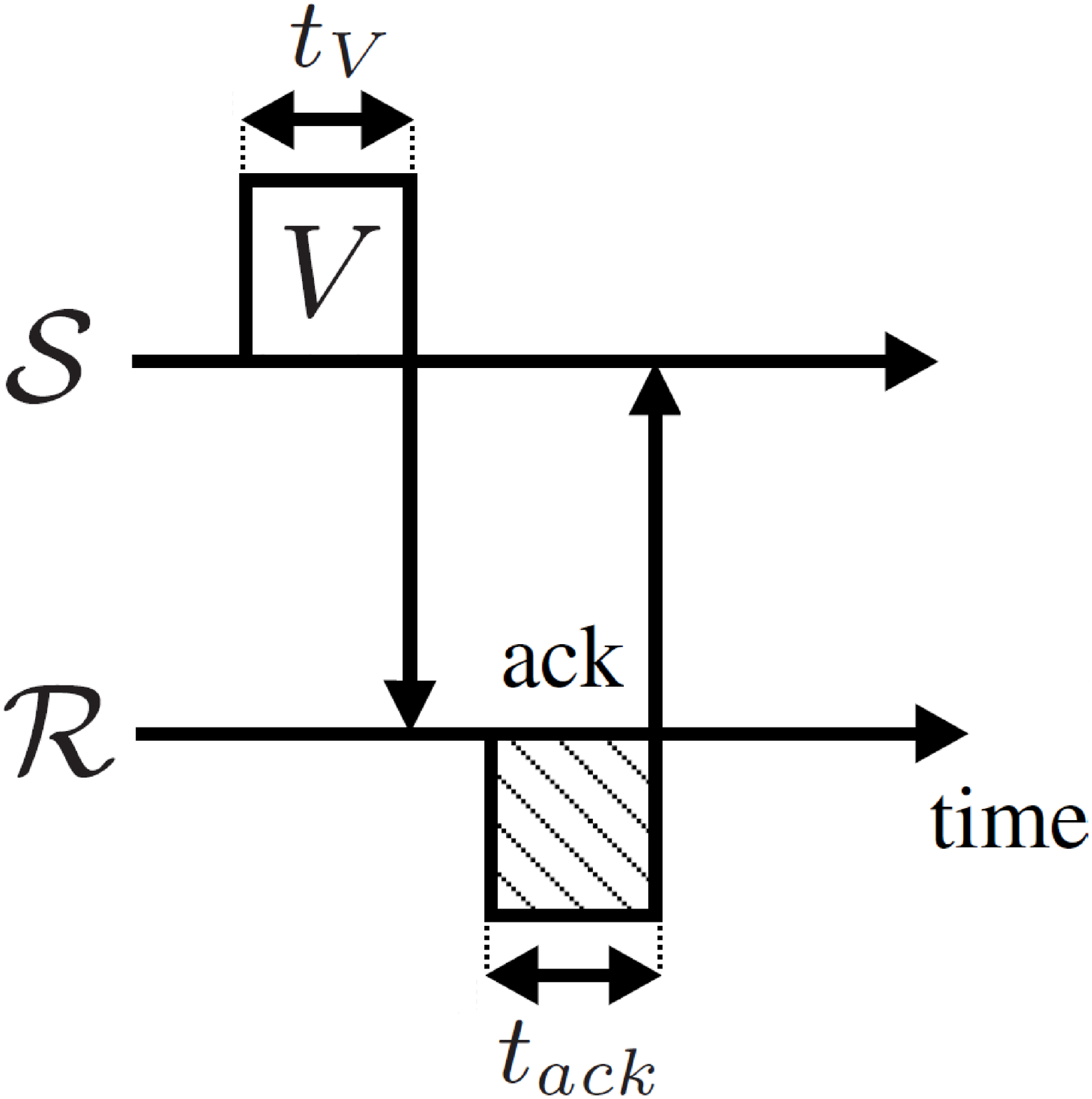}	
			\label{fig:drawing_ACK2}		
		} 	
		\subfigure[Jam-2]{
			\includegraphics[width=0.2675\textwidth]{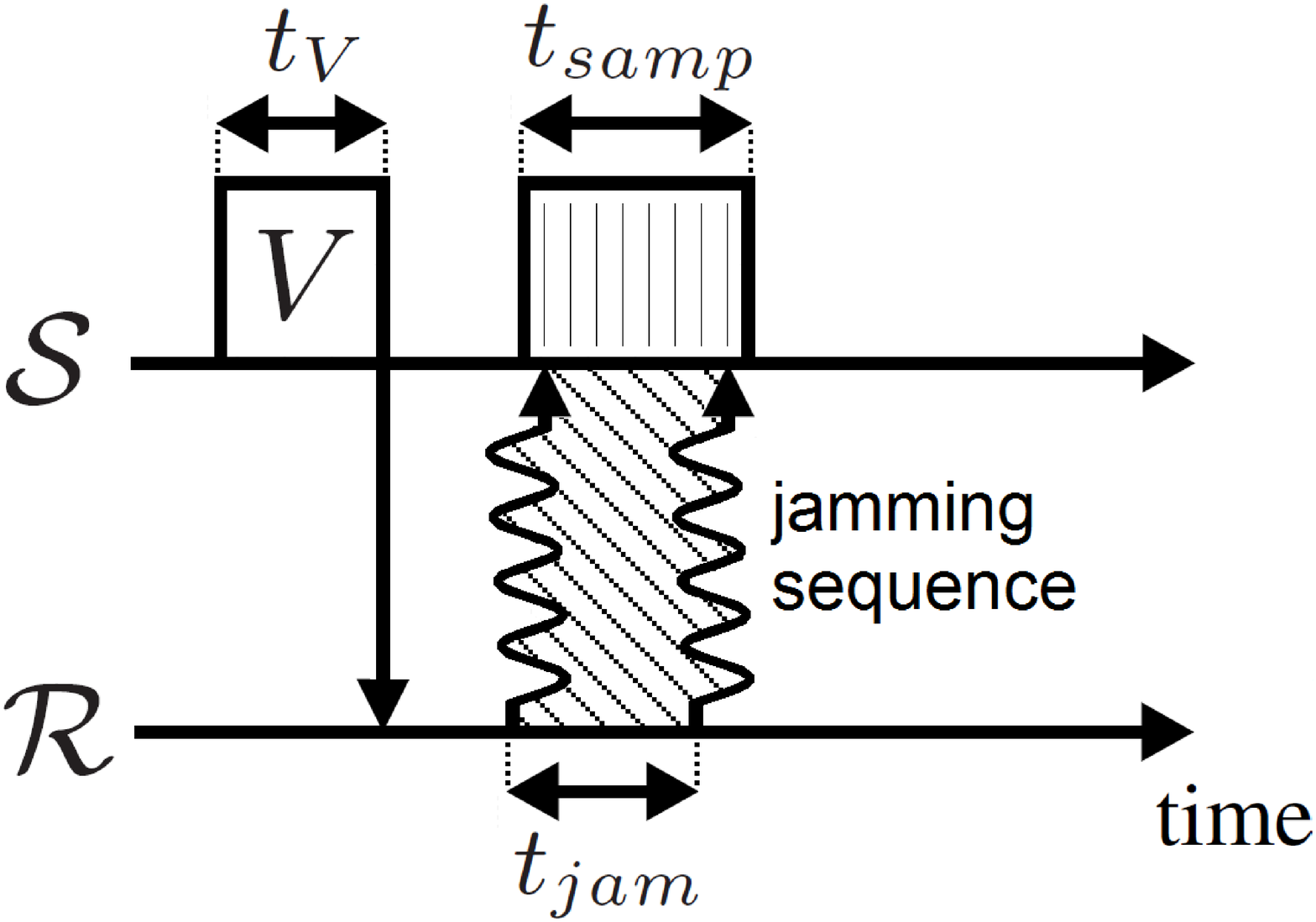}
			\label{fig:drawing_JAM2}		
		} 	
		\subfigure[Jam-3]{
			\includegraphics[width=0.2675\textwidth]{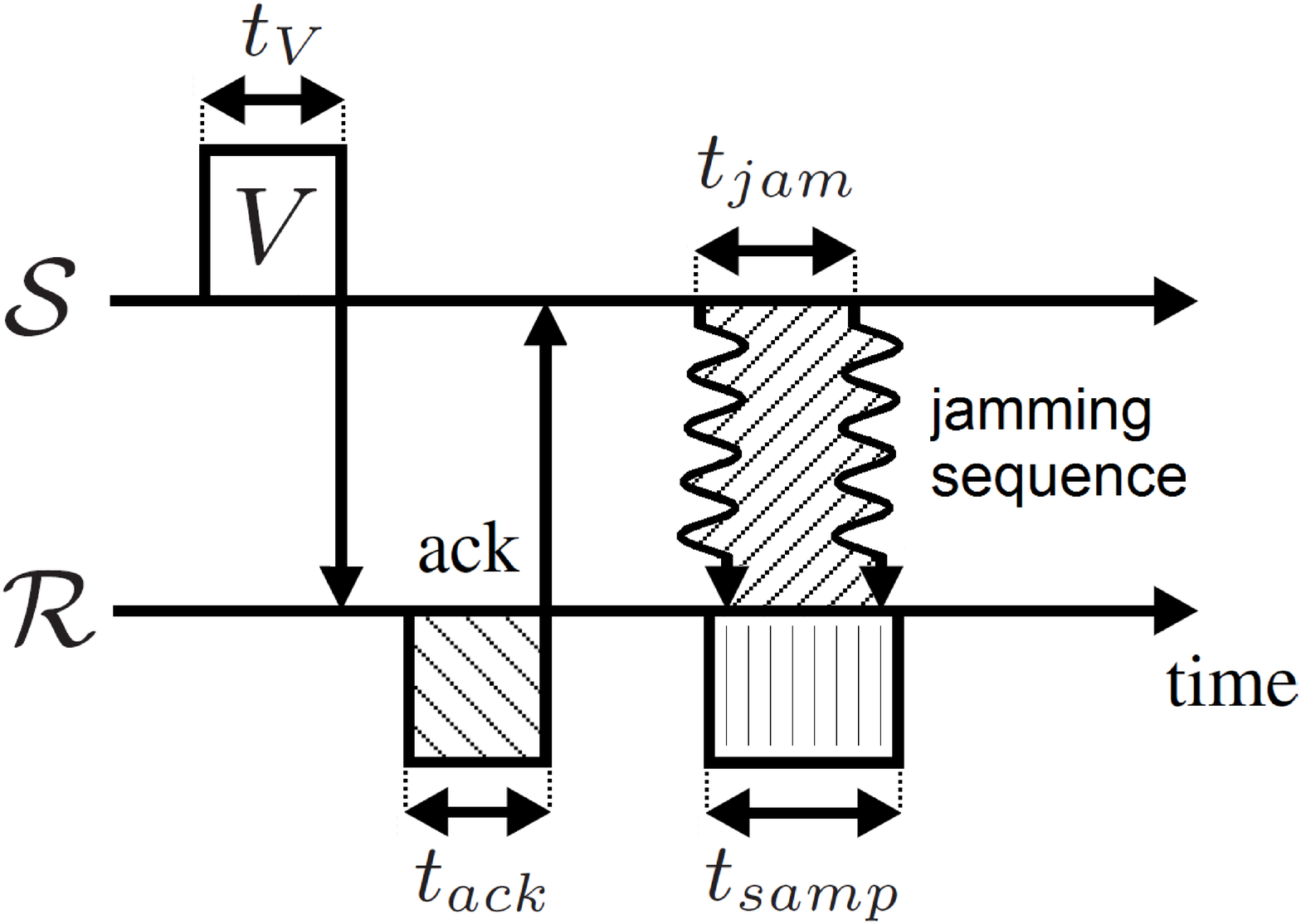}
			\label{fig:drawing_JAM3}		
		} 			
		\vspace{-3mm}
		\caption{Illustration of agreement protocols.}
		\vspace{-5mm}
 		\label{fig:drawings_2}
	\end{center}
\end{figure*}

\subsection{Jam-2: 2-Way Handshake with Jamming} \label{subsec:jam_2}

As highlighted in Section~\ref{sec_problem}, due to the constraints of wireless
sensor networks we need to minimize the amount of exchanged packets $n$. For
$n=2$ we obtain a two-way handshake protocol shown in
Figure~\ref{fig:drawing_ACK2} where node
$\mathcal{S}$ initiates the transmission and sends the value $V$ to node
$\mathcal{R}$, and node $\mathcal{R}$ acknowledges the receipt of $V$ with a new
transmission towards $\mathcal{S}$. We refer to this protocol as \emph{Ack-2}.

We call \emph{Jam-2} the two-way handshake in which the acknowledgment is sent
in the form of a jamming sequence (Figure~\ref{fig:drawing_JAM2}). Node
$\mathcal{S}$ initiates the exchange and sends the information $V$ towards a
receiver $\mathcal{R}$ (Figure~\ref{fig:drawing_JAM2}). If the first message is
successfully received, $\mathcal{R}$ transmits a jamming signal for a period
$t_{jam}$. Meanwhile, $\mathcal{S}$ carries out a fast RSSI sampling for a
period $t_{samp} \le t_{jam}$ that is synchronized in such a way that the fast
RSSI sampling is carried out while the jamming signal is on the air. The first
message, i.e., the one containing the information $V$, is used as the
synchronization signal. For our purposes, this is sufficient since clock drift
is insignificant at timescales of a few milliseconds. Hence, for simplicity, in
the rest of the paper we assume $t_{jam} = t_{samp}$.

Denoting $r_{noise}$ as the maximum RSSI noise floor value measured in the
absence of interference,
and $\{x_1, x_2, \ldots, x_n\}$ as the sequence of RSSI values sampled during
$t_{samp}$, we define the binary sequence $\{X_1, X_2, \ldots, X_n\}$ as
follows: if $x_i \le r_{noise}$, then $X_i=1$, else $X_i=0$. If $\sum_{i=1}^{n} X_i = 0$,
the initiator $\mathcal{S}$ assumes that a jamming sequence was transmitted by
node $\mathcal{R}$.

Note that the received signal strength $r_{jamming}$ of the jamming
sequence has to be higher than $r_{noise}$. Therefore, the selection of
$r_{noise}$ is important, especially for low-quality links, where $r_{jamming}$
will have a value very close to $r_{noise}$.

In order for Jam-2 to be successful (positive agreement), two
conditions must be satisfied: (i) the packet from node $\mathcal{S}$
to node $\mathcal{R}$ must be successfully decoded and (ii) node
$\mathcal{S}$ must correctly detect the presence of the jamming signal
and not confuse it with any other external interference signal.

Condition (i) depends on the interplay of $IDLE_{i}$
(length of idle periods in the interference signal) and the duration
$t_{V}$ of the transmission of the packet containing $V$. A necessary
condition for receiving this message is $t_{V}<\max\{IDLE_{i}\}$.
If $\mathcal{R}$ cannot successfully decode the packet, $\mathcal{R}$
does not transmit any jamming sequence.
The success of condition (ii) depends on the interplay of $BUSY_{i}$
(length of busy periods in the interference signal) and the duration
$t_{jam}$ of the jamming signal. As discussed in
Section~\ref{subsec:detecting}, a necessary condition for correctly
detecting the jamming signal is $t_{jam} > \max\{BUSY_{i}\}$.

Increasing the duration $t_{jam}$ of the jamming signal leads
to longer delays and a higher energy consumption. Having $t_{jam} <
\max\{BUSY_{i}\}$ may imply that $X_i$ is always $0$, and
$\mathcal{S}$ concludes that a jamming sequence was transmitted even
though it was not. This scenario may lead to disagreement that we call
false positive agreement when $\mathcal{R}$ does not receive the
packet containing $V$ and does not transmit any jamming sequence, but
the interference lasts for a period longer than $t_{jam}$. \\

\vspace{-2.5mm}
In order for Ack-2 to be successful (positive agreement), both the packet from
$\mathcal{S}$ to $\mathcal{R}$ and the acknowledgment packet from $\mathcal{R}$
to $\mathcal{S}$ must be successfully decoded. The success of the exchange
depends on the interplay of $IDLE_{i}$ and the transmission delays $(t_{V}$ and
$t_{ack}$, see also Section~\ref{subsec:implementation}).



\vspace{+1.5mm}
\subsection{Jam-3: 3-Way Handshake with Jamming}  \label{subsec:jam_3}

When using Jam-2, node $\mathcal{S}$ can confuse external interference
resulting in RSSI values greater than $r_{noise}$ with the jamming sequence
transmitted by node $\mathcal{R}$. Having an approximate knowledge of the signal
strength $r_{jamming}$ of the jamming sequence would enable $\mathcal{S}$ to
filter out RSSI values $r_{noise} < x_i < r_{jamming}$ that lead to $X_i=1$
in Jam-2, therefore reducing the probability of a false positive agreement.

Therefore we consider a 3-way handshake, where $\mathcal{R}$ can exploit
knowledge of the RSSI of the first received packet to enhance the detection of a
later jamming sequence, assuming that the two should have similar received
signal strength. The rest of this section describes Jam-3, a three-way handshake
protocol where the last acknowledgement is implemented with jamming as shown in
Figure~\ref{fig:drawing_JAM3}.


Node $\mathcal{S}$ initiates the exchange and sends the information
$V$ towards a receiver $\mathcal{R}$. If the first message is received
successfully, $\mathcal{R}$ sends an acknowledgment packet back to
$\mathcal{S}$. If $\mathcal{S}$ receives the ACK, it transmits a
jamming signal for a period $t_{jam}$. Meanwhile, $\mathcal{R}$
carries out a fast RSSI sampling for a period $t_{samp} \le
t_{jam}$ (Figure~\ref{fig:drawing_JAM3}). If the receiver detects the
jamming signal from the sender it deems the exchange as successful;
otherwise, $V$ is discarded.

The advantage of this approach compared to Jam-2 is that $\mathcal{R}$ knows the
received signal strength of the first packet $r_{s}$. Under the
hypothesis that the jamming signal has a reasonably similar strength as $r_{s}$,
$\mathcal{R}$ filters out RSSI values $r_{i}<r_{s}$ that must have been
generated by external interference.

Hence, denoting $\{x_1, x_2, \ldots, x_n\}$ as the sequence of RSSI values
sampled during $t_{samp}$, we define the binary sequence $\{X_1, X_2, \ldots, X_n\}$
as follows: if $x_i < (r_{s}-\Delta_{r})$, then $X_i=1$, else $X_i=0$, where
$\Delta_{r}$ is a threshold to account for slight variations in the signal strength.
If $\sum_{i=1}^{n} X_i = 0$, $\mathcal{R}$ assumes that a jamming
sequence was transmitted by $\mathcal{S}$.

Due to the inaccuracy of low-power radios, a tolerance margin $\Delta_{r}$ must
be provided, and its role is crucial. If $(r_{s}-\Delta_{r}) < r_{noise}$, then
the algorithm assumes $r_{s} := r_{noise}$, resulting in the same behavior as in
Jam-2.
Compared to Jam-2, Jam-3 can provide a higher accuracy because it exploits the
information provided by $r_{s}$. As with Jam-2, false positive agreements can be
removed if $t_{jam} > \max\{BUSY_{i}\}$ at the expense of a higher energy
consumption.


\vspace{+1.5mm}
\subsection{Jam-X Implementation} \label{subsec:implementation}
We implement Jam-X on the Tmote Sky platform that features a CC2420
radio~\cite{chipconCC2420} and an MSP430 microprocessor. Our
implementation, based on Contiki~\cite{dunkels04contiki}, uses two 
main building blocks, i.e., the generation of the jamming sequence and
the high-frequency RSSI sampling.

The generation of the jamming sequence uses the CC2420 transmit
test modes as described in Section~\ref{subsec:jamming_generation}. 
This specific feature of the CC2420 transceiver is also available
on other nodes such as the Ember EM2420 and the CC2520~\cite{boano11jamlab}.
Similar to Boano et al.~\cite{boano11jamlab} we implement the
high-frequency RSSI sampling by boosting the CPU speed and
optimizing the SPI operations. This way, we 
obtain one RSSI sample approximately every 20 $\mu$s. 
The achieved sampling rate of about 50 KHz does not capture the
transmissions from all other devices operating in the same
frequency band such as IEEE~802.11n. Nevertheless,
we can still identify most of the idle instants between Wi-Fi
transmissions which allows us to distinguish the jamming
sequence from external interference as discussed in Section \ref{subsec:detecting}.

For the experiments we use NULLMAC, a MAC layer that just forwards
packets to the upper or lower protocol layer and does not perform any
duty cycling. We chose NULLMAC in order to obtain results that are
independent of specific MAC features and parameters.
%
%
Before sending a message, a Clear Channel Assessment (CCA) is performed to
minimize the chances that the packet is destroyed by interference. For the
initial handshake message containing $V$, $\mathcal{S}$ waits until the CCA
signals an idle medium. At this point, the initiator of the handshake transmits
the first packet.

Before transmitting subsequent ACK packets (but not before jamming), a CCA check
is performed once and the packet is only transmitted if the medium is idle. The
reason for not waiting until the channel clears is that this would introduce
unbounded delays such that it could not be decided if the acknowledgement is
lost or just arbitrarily delayed. \\


\vspace{+3mm}

%% file: evaluations_unicast_2.tex
\vspace{-1.5mm}
\section{Unicast Evaluation} \label{sec:evaluation_unicast}
\vspace{-0.5mm}
In this section, we evaluate the performance of Jam-2 and Jam-3 by
comparing it with the performance of packet-based handshakes for
unicast agreement.

\vspace{-0.5mm}
\subsection{Experimental Setup} \label{subsec:experimental_setup}
\vspace{-0.5mm}
In order to evaluate the performance of Jam-X, we carry out experiments in two
small-scale indoor testbeds deployed in office environments with USB-powered
motes. In the first testbed, we use JamLab, a tool for controlled interference
generation~\cite{boano11jamlab} to evaluate the impact of interference in a
realistic and repeatable fashion. In JamLab, interference is either replayed
from trace files that contain RSSI values recorded under interference, or from
models of specific devices~\cite{boano11jamlab}. In particular, we use
JamLab to emulate the interference patterns produced by microwave ovens, by
Bluetooth, and by Wi-Fi devices. In the latter case, the interference emulates a
continuous file transfer. To avoid additional interference as much as possible,
we carry out the experiments in this testbed during the night, when Wi-Fi
activity in the office building is lowest.
In the second testbed, we do not use JamLab, but we deliberately choose an
802.15.4 channel affected by interference, namely channel~18. On channel~18
there is Wi-Fi traffic and sometimes also interference from microwave ovens in a
nearby kitchen.

For the experiments, we use two motes $\mathcal{S}$ and $\mathcal{R}$. Node
$\mathcal{S}$ always initiates the handshake, and transmits a data packet $V$
composed of a 4-byte sequence number and one additional byte containing the
transmission power used. For each handshake, we select a random transmission
power between -25~dBm and 0~dBm. $\mathcal{R}$ replies to the message using the
transmission power contained in the packet, i.e., the same one used by
$\mathcal{S}$. By using different transmission powers, we create different types
of links for each handshake. Each packet is sent after a random interval in the
order of tens of milliseconds, and nodes remain on the same channel for the
whole duration of the experiment. Each experiment consists of several hundred
thousand handshakes.


\begin{figure*}[t!]
	\begin{center}	
		\subfigure[Bluetooth]{
			\includegraphics[width=0.30\textwidth]{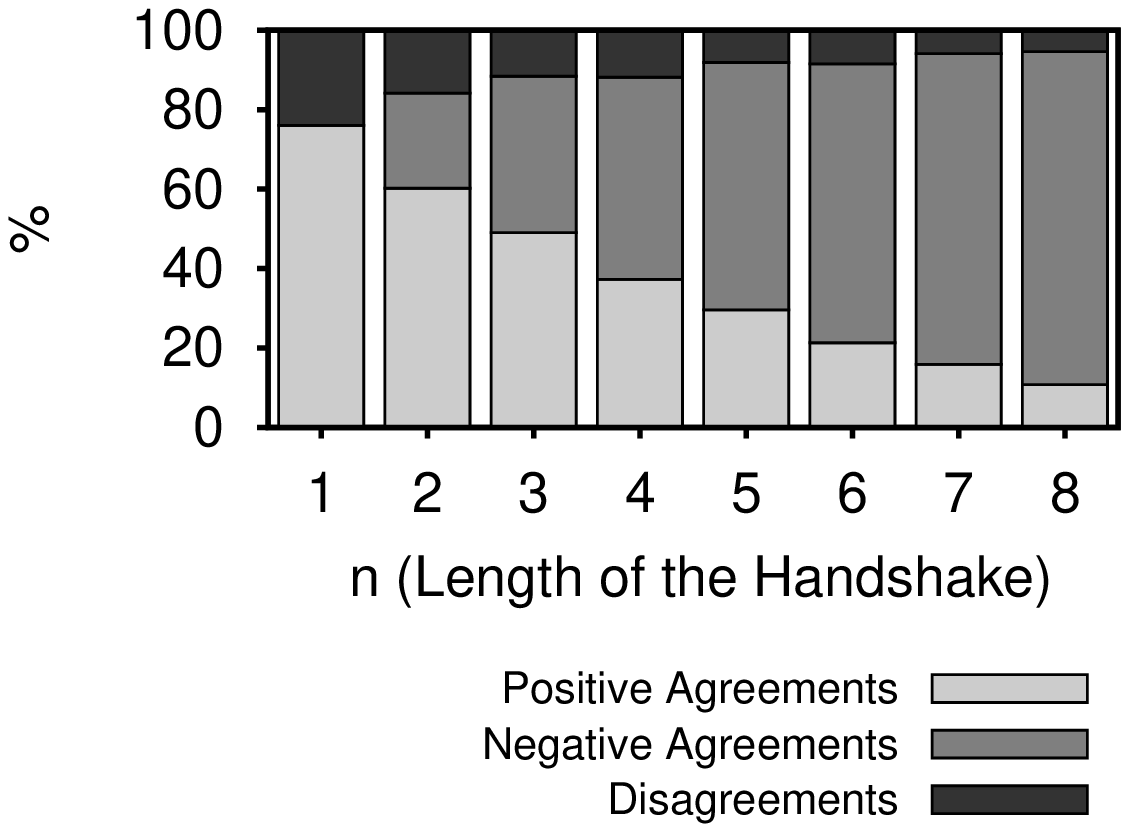}
			\label{fig:ACK_handshake_bluetooth}
		}	
		\subfigure[Microwave oven]{
			\includegraphics[width=0.30\textwidth]{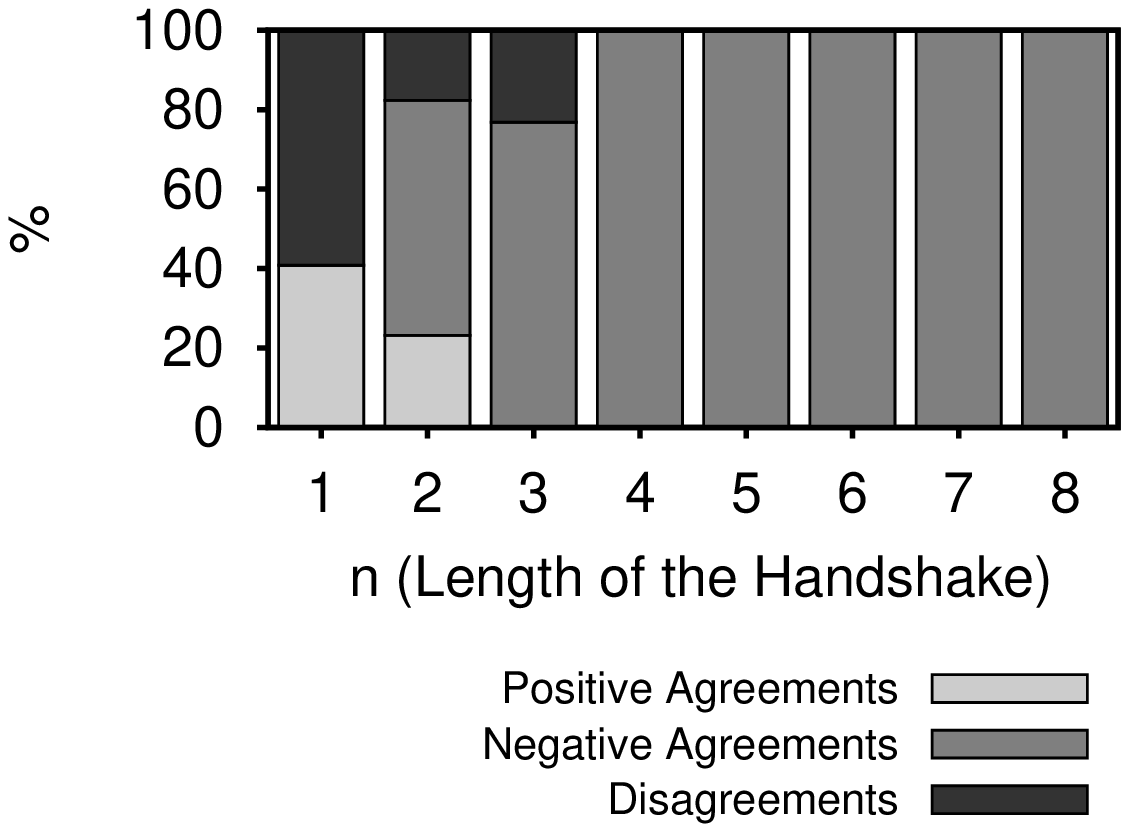}
			\label{fig:ACK_handshake_microwave}		
		}			
		\subfigure[Heavy Wi-Fi]{
			\includegraphics[width=0.30\textwidth]{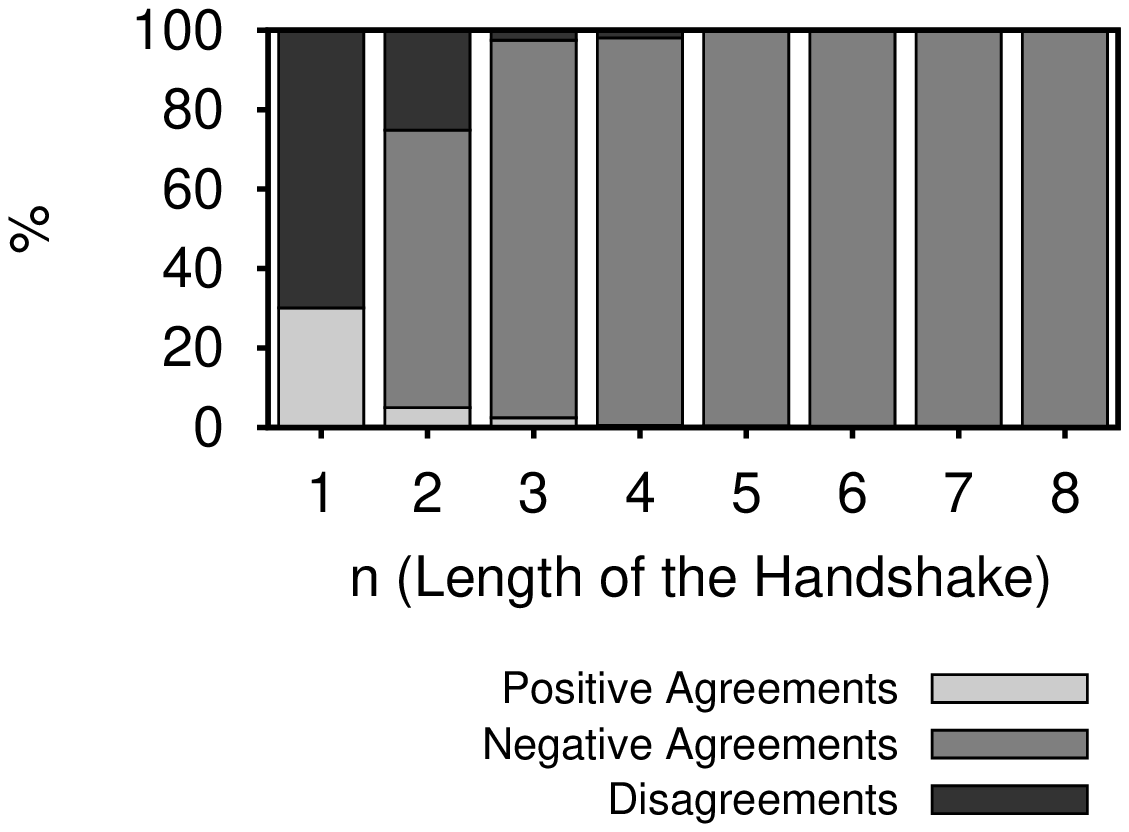}
			\label{fig:ACK_handshake_wifi}
		}		
		\vspace{-3.5mm}				
		\caption{Performance of a n-way handshake under different types of interference.
		The initiator does not wait for the channel to be free before
		transmitting the first packet.}
		\vspace{-7.5mm}	
 		\label{fig:ACK_handshake_interference}
	\end{center}
\end{figure*}

\begin{figure*}[t!]
	\begin{center}	
		\subfigure[Bluetooth]{
			\includegraphics[width=0.30\textwidth]{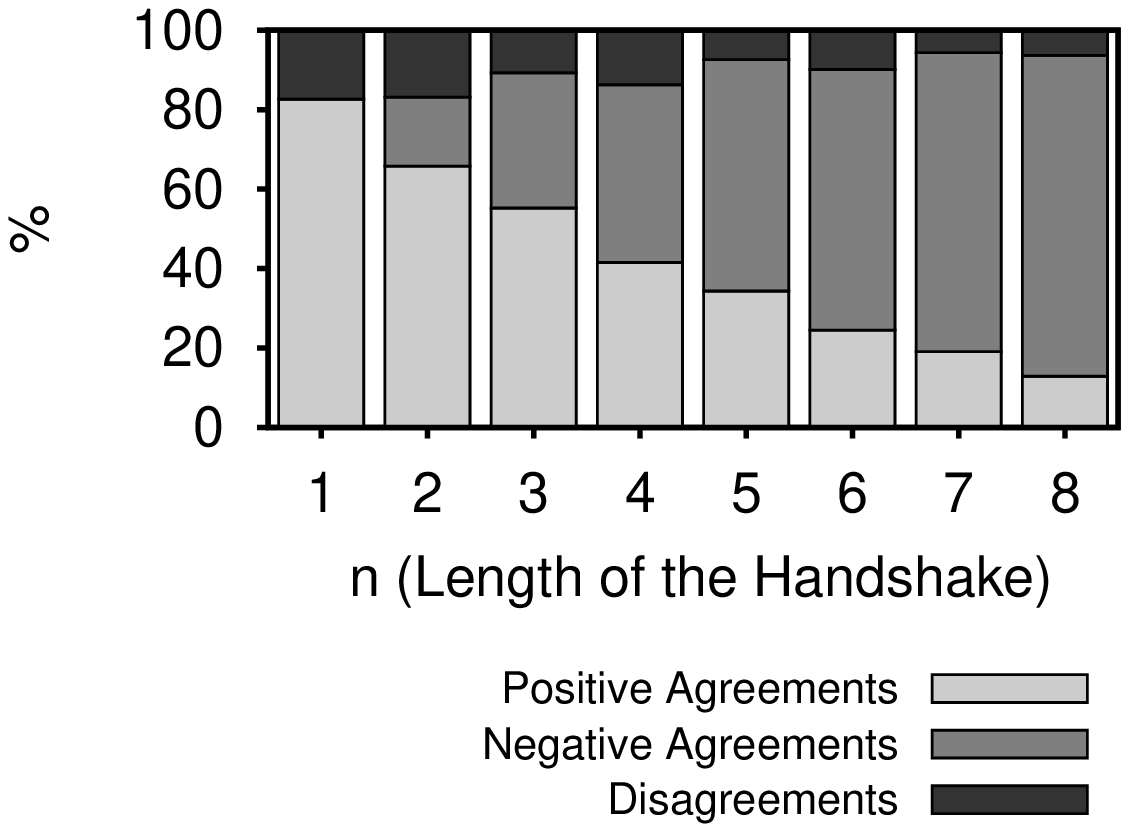}
			\label{fig:CCA_ACK_handshake_bluetooth}
		}	
		\subfigure[Microwave oven]{
			\includegraphics[width=0.30\textwidth]{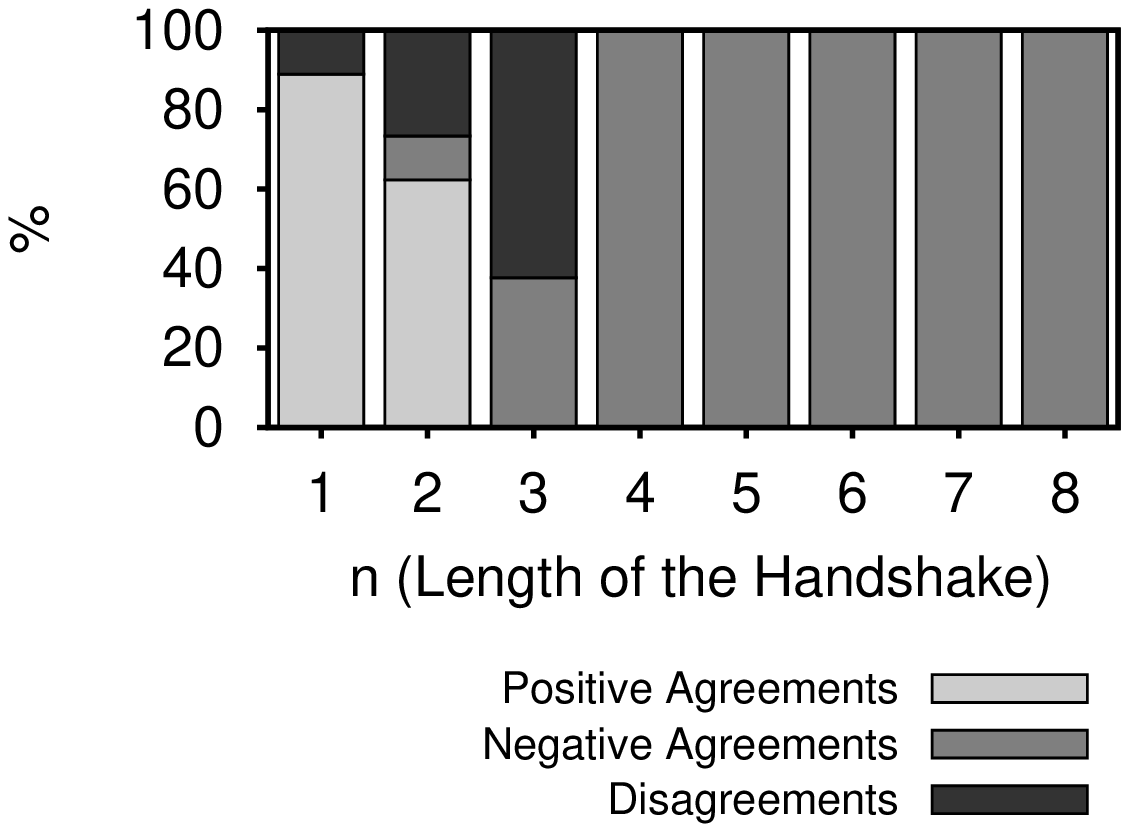}
			\label{fig:CCA_ACK_handshake_microwave}		
		}			
		\subfigure[Heavy Wi-Fi]{
			\includegraphics[width=0.30\textwidth]{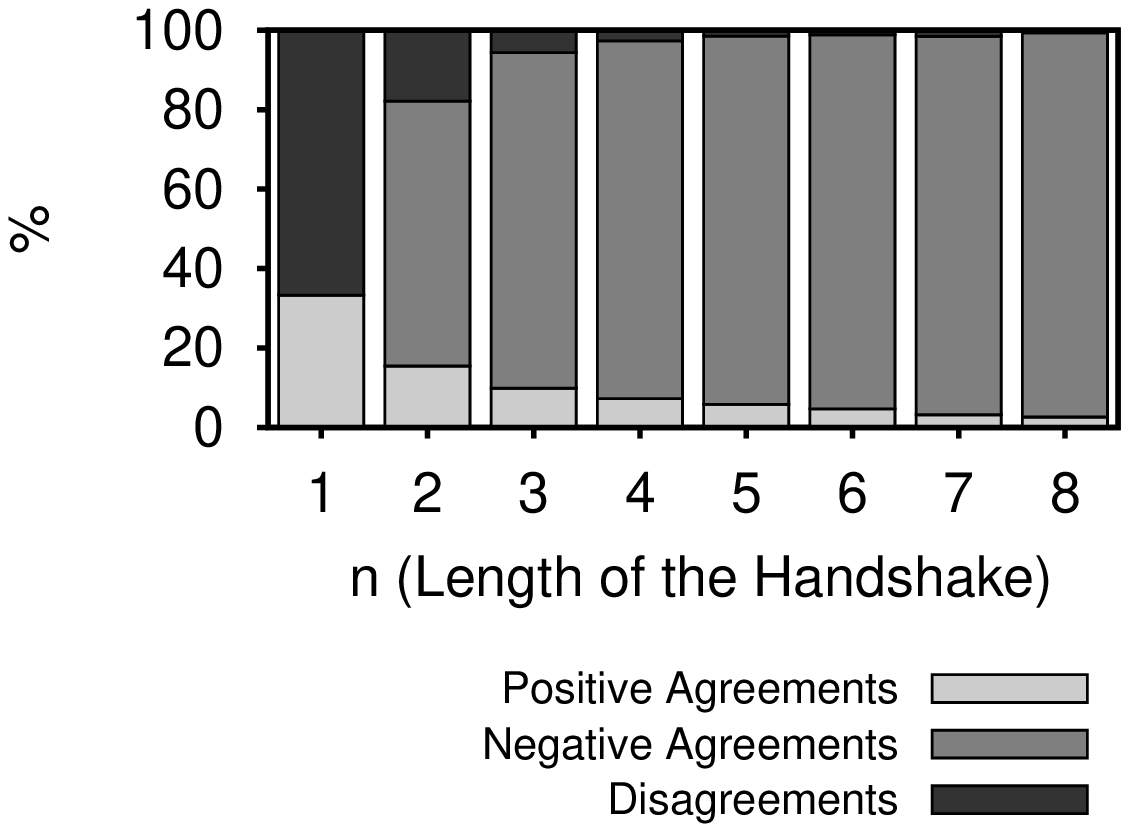}
			\label{fig:CCA_ACK_handshake_wifi}
		}		
		\vspace{-3.5mm}		
		\caption{Performance of a n-way handshake under different types of interference.
		The initiator waits for the channel to be free before transmitting the first
		packet.}
		\vspace{-6mm}	
 		\label{fig:CCA_ACK_handshake_interference}
	\end{center}
\end{figure*}

\subsection{Generic n-way Handshake} \label{subsec:evaluation_unicast_n_way_handshake}

Using JamLab, we analyze the performance of a generic n-way handshake under
different interference patterns. Fig.~\ref{fig:ACK_handshake_interference}
and~\ref{fig:CCA_ACK_handshake_interference} show the percentage of
positive/negative agreements and disagreements under interference for n-way
handshakes, i.e., the protocol shown in Fig.~\ref{fig:drawing_handshake}, with
up to $n=8$ packets.

The results match very well the trends of the theoretical results shown
Figure~\ref{fig:probabilities_3d}. When $n$ increases, the probability of both
disagreements and positive agreements decreases. The probability of
negative agreements grows with $n$.

As expected, under microwave oven interference (that has a periodic pattern
with idle and busy durations of $\approx$ 10 ms), positive agreement cannot be
reached when $n \ge 3$. This is because the duration of the handshake is longer
than the idle period, hence interference will hit at least one of the
packets.

We also investigate the impact of performing CCA before transmitting the
first packet as described in Section~\ref{subsec:implementation}.
Figure~\ref{fig:ACK_handshake_interference} is without CCA, while
Figure~\ref{fig:CCA_ACK_handshake_interference} is with CCA. Generally speaking,
performing CCA improves the performance, but not significantly. In the case of
the microwave oven, performing the CCA effectively searches for the begin of an
idle period, thus maximizing the time available to complete the handshake
before interference starts again. Therefore, in the remaining experiments
we always perform CCA before sending the first packet.

\subsection{2-way Handshakes: Ack-2 and Jam-2} \label{subsec:evaluation_unicast_2_way_handshake}

\begin{figure*}[t!]
	\begin{center}	
		\subfigure[Bluetooth]{
			\includegraphics[width=0.30\textwidth]{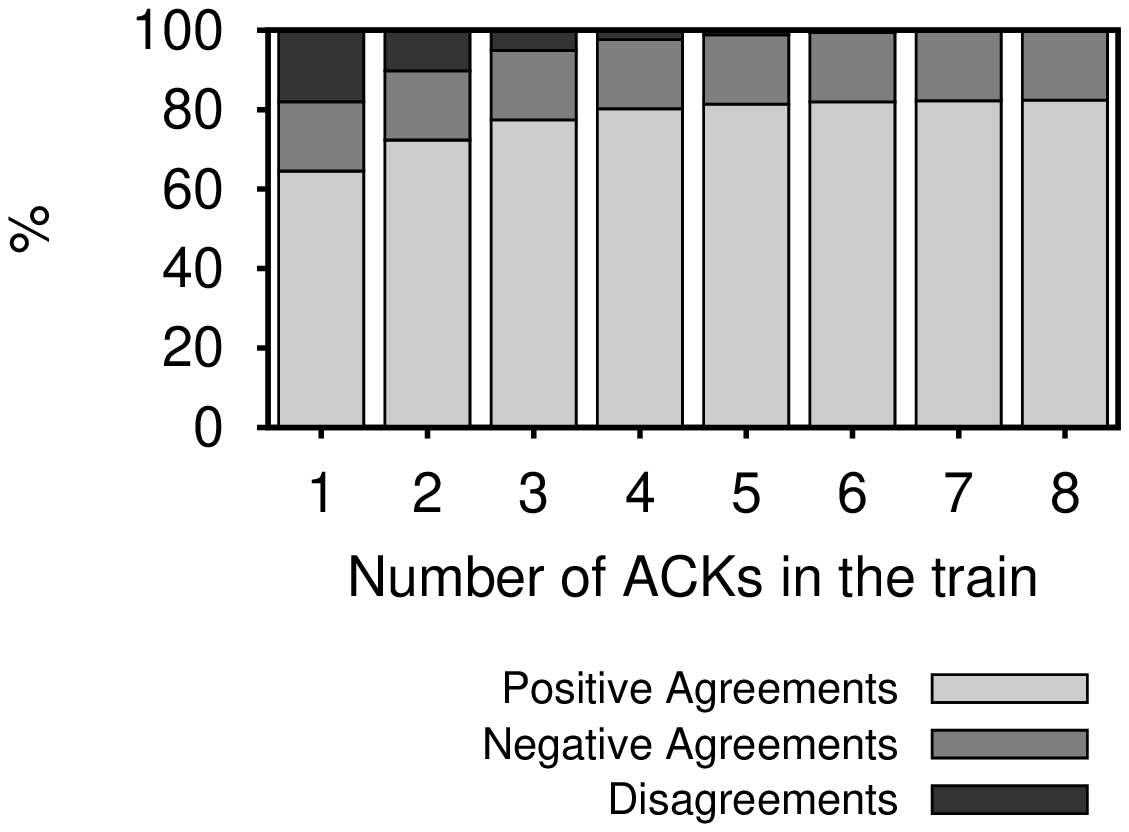}
			\label{fig:CCA_ACK2_bluetooth}
		}	
		\subfigure[Microwave oven]{
			\includegraphics[width=0.30\textwidth]{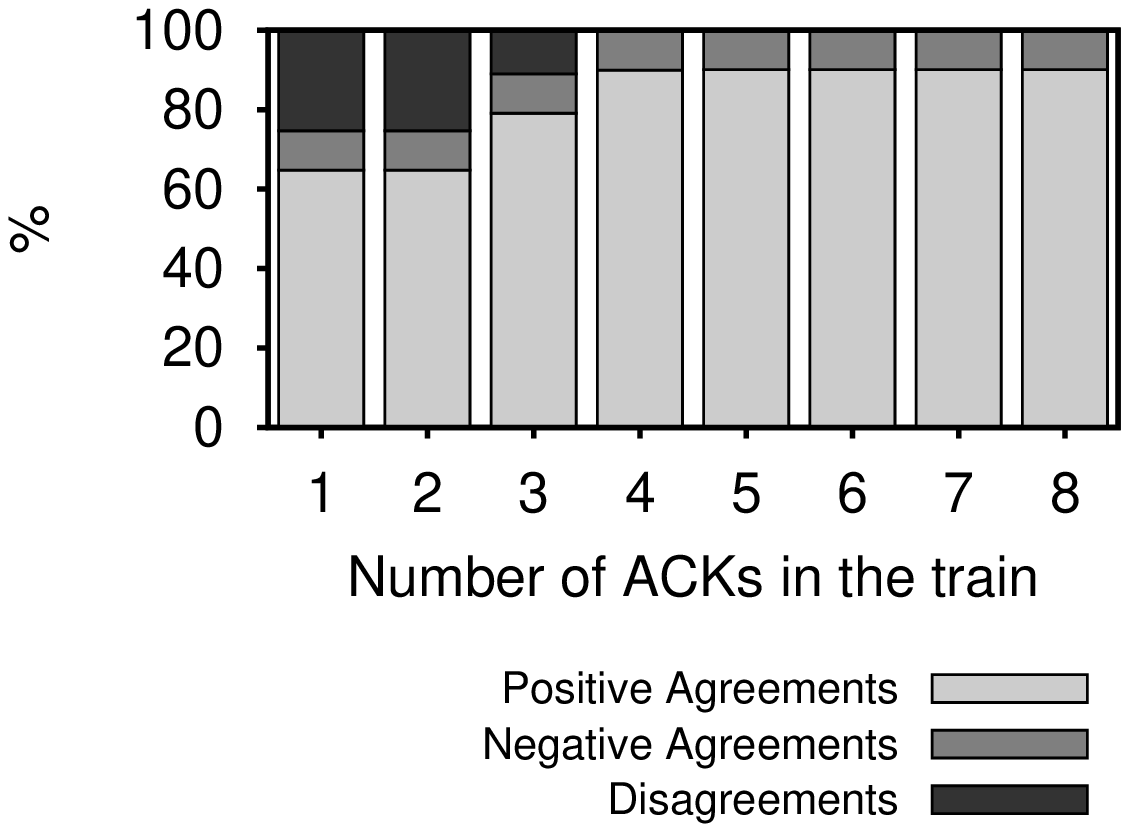}
			\label{fig:CCA_ACK2_microwave}		
		}			
		\subfigure[Heavy Wi-Fi]{
			\includegraphics[width=0.30\textwidth]{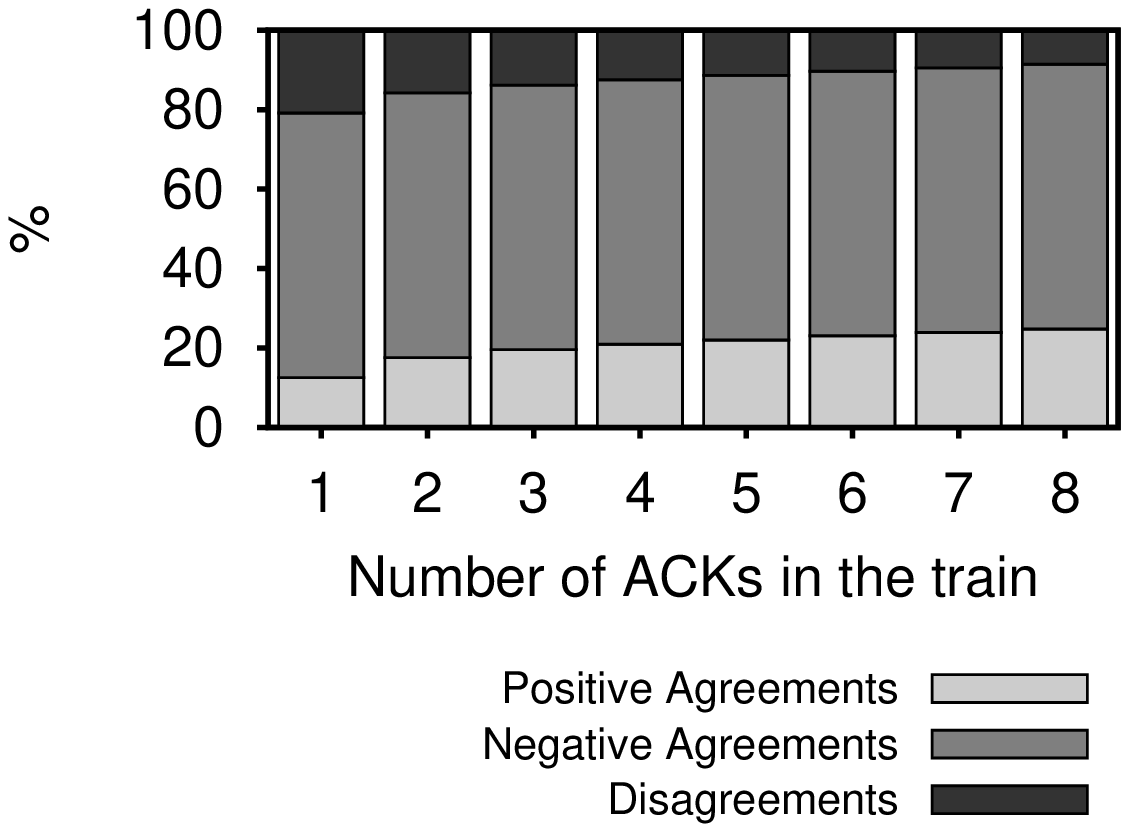}
			\label{fig:CCA_ACK2_wifi}
		}			
		\vspace{-3.5mm}		
		\caption{Trains of acknowledgment packets improve the performance of the
		two-way handshake Ack-2.}
		\vspace{-7mm}	
 		\label{fig:CCA_ACK2_interference}
	\end{center}
\end{figure*}

We now compare the performance of 2-way handshakes based on packet
transmissions (Ack-2) and jamming sequences (Jam-2).
Figure~\ref{fig:CCA_ACK2_interference} shows the performance of Ack-2
under interference. As discussed in Section~\ref{subsec:agreement_wsn}
we investigate the case where a train of $T$ acknowledgement
packets is sent to increase the probability of successful reception. The
case in which $T=1$ corresponds to the 2-way handshake depicted in
Figure~\ref{fig:drawing_ACK2}.

It is important to highlight two aspects related to sending a train of
acks. Firstly, having $T>1$ improves the probability of positive
agreements.  Secondly, sending a train of $k$ packets is much faster
than performing a $k$-way handshake as the packet can be loaded once
into the radio and then transmitted several times, thus avoiding the
overhead for receiving, decoding, extracting, analyzing, processing
the packet and generating the response.  Due to this reason, we can
send more repeated ACKs during the idle time of the microwave than
we can send handshake messages.


\begin{figure*}[t!]
	\begin{center}	
		\subfigure[Bluetooth]{
			\includegraphics[width=0.30\textwidth]{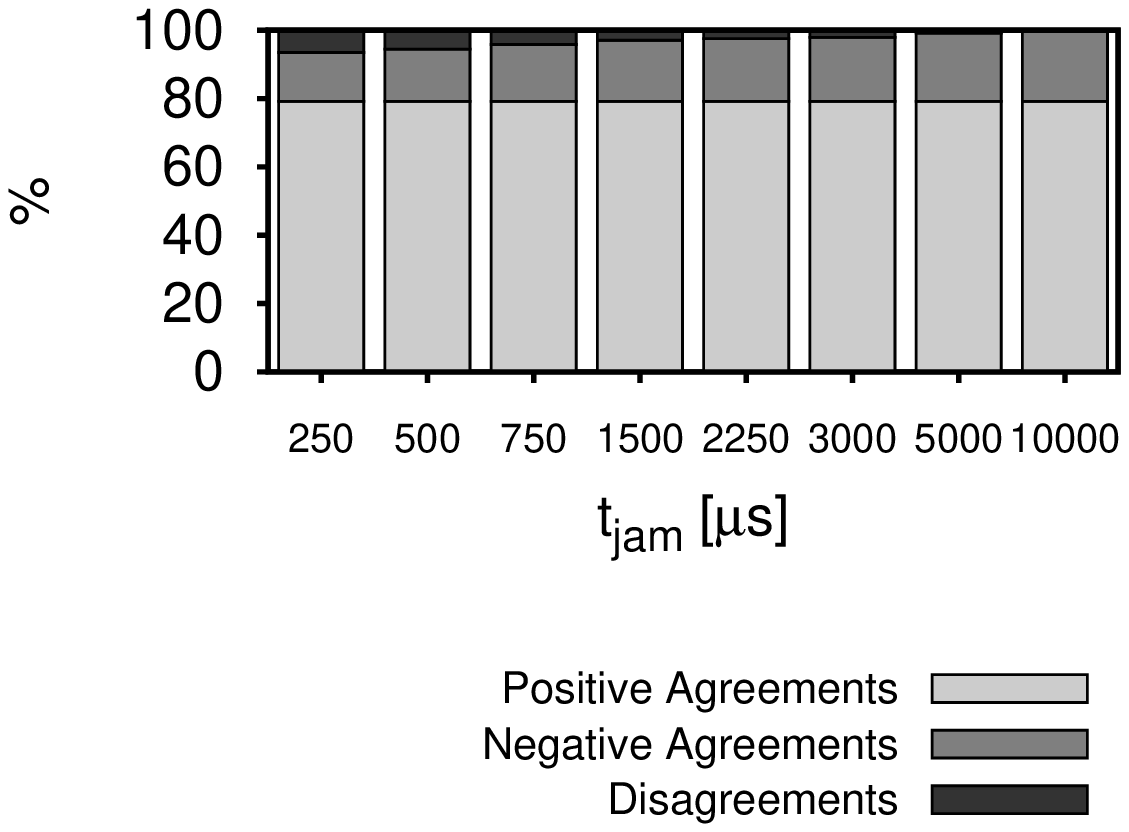}
			\label{fig:CCA_JAM2_bluetooth}
		}	
		\subfigure[Microwave oven]{
			\includegraphics[width=0.30\textwidth]{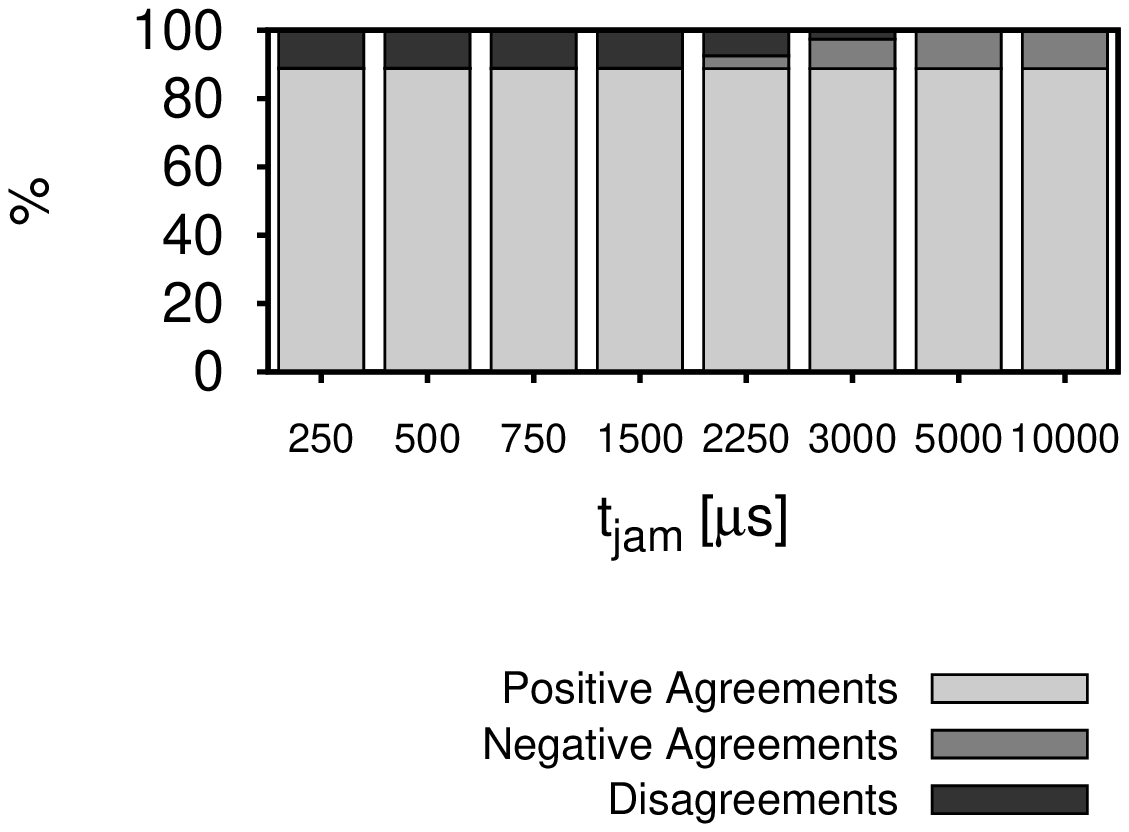}
			\label{fig:CCA_JAM2_microwave}		
		}			
		\subfigure[Heavy Wi-Fi]{
			\includegraphics[width=0.30\textwidth]{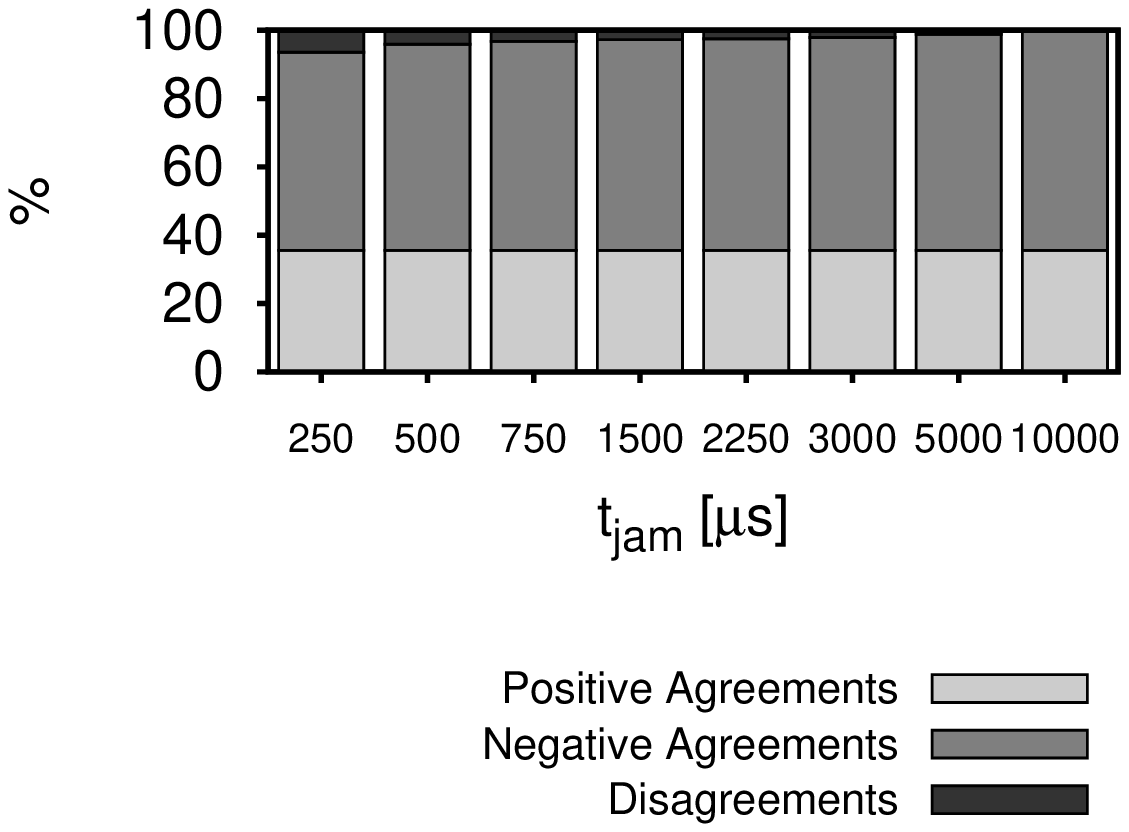}
			\label{fig:CCA_JAM2_wifi}
		}		
		\vspace{-3.5mm}
		\caption{Performance of Jam-2: compared to Ack-2
		(Figure~\ref{fig:CCA_ACK2_interference}), Jam-2 shows better performance
		for all interfering sources, as it maximizes the amount of positive
		agreements and minimizes the amount of disagreements.}
		\vspace{-6mm}	
 		\label{fig:CCA_JAM2_interference}
	\end{center}
\end{figure*}

Figure~\ref{fig:CCA_JAM2_interference} shows the probability of positive/negative
agreements and disagreements under interference for Jam-2. After receiving the
first packet, $\mathcal{R}$ transmits a jamming sequence of length $t_{jam}$.
Obviously, the longer the jamming sequence lasts, the higher the probability of
successful agreement. 

Observe the following two important aspects. Firstly, the probability
of disagreements with Jam-2 is much lower than that of Ack-2 even for
short jamming periods $t_{jam}$. Secondly, in Ack-2 the amount of
negative agreements is basically independent on $T$, and with
increasing $T$ an increasing fraction of the disagreements is replaced
by positive agreements. Instead, with Jam-2, the amount of positive
agreements remains constant and reaches the maximum already with short
$t_{jam}$. With increasing $t_{jam}$, an increasing fraction of the
disagreements is replaced by negative agreements. This is because in
Jam-2 the detection of a jamming signal is \emph{very} reliable, and
the disagreements basically occur only when the first packet is lost --
resulting in negative agreement.

%

We now investigate the energy consumption and time to completion of Jam-2 and
Ack-2. For this purpose, we first measure the transmission delay (i.e., the time
in which the message is actually over the air) as a measure of transmit power
consumption. Secondly, we measure the time that is actually required by the
operating system to complete the \emph{unicast\_send} command. To achieve this
goal, we use Contiki's software-based power
profiler~\cite{dunkels07softwarebased}. The second value is larger than the
first one, because it includes the processing of the packet and its loading into
the radio buffer before the packet actually gets transmitted over the air.

\begin{figure*}[t!]
	\begin{center}
		\subfigure[Time required for transmissions]{
			\includegraphics[width=0.32\textwidth]{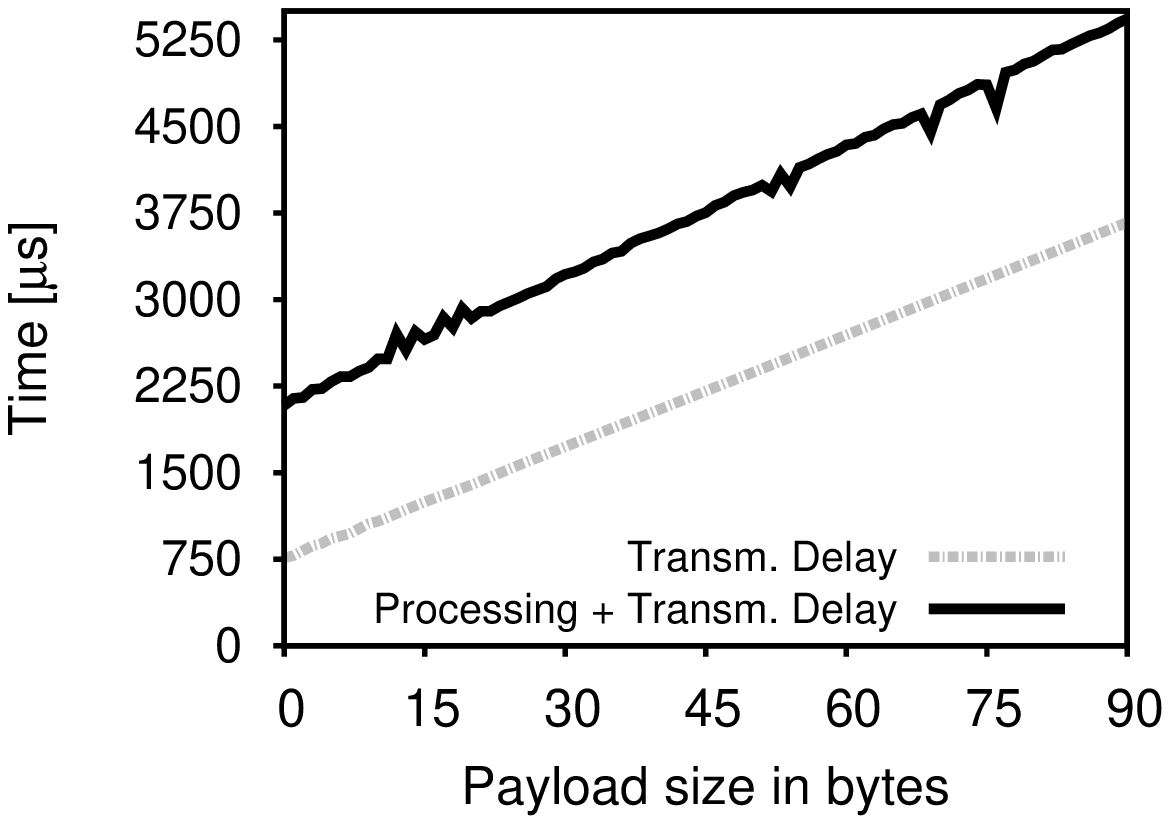}
			\label{fig:tx_delay}
		}
		\subfigure[Disagreements as function of time]{
			\includegraphics[width=0.32\textwidth]{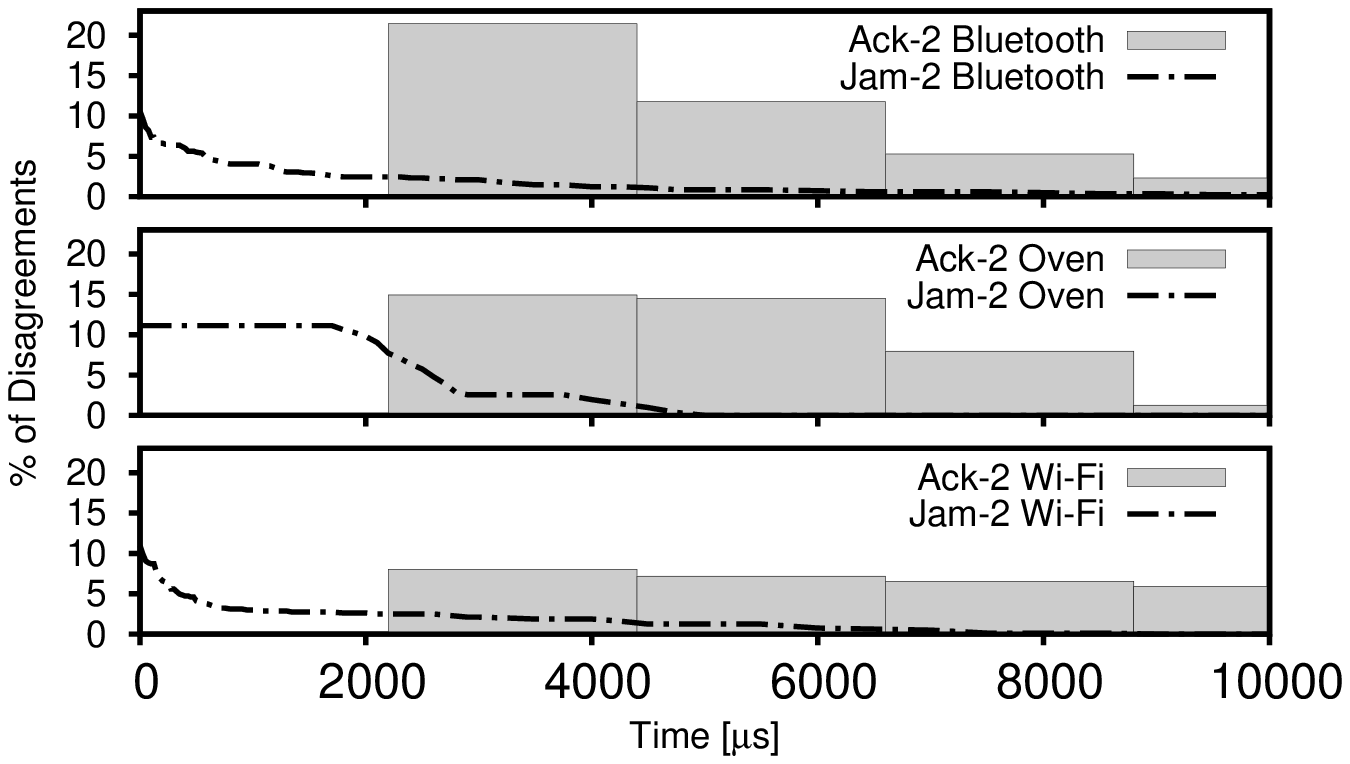}
			\label{fig:JAM2_ACK2_disagreements_time}		
		}		
		\subfigure[Disagreements as function of energy]{
			\includegraphics[width=0.32\textwidth]{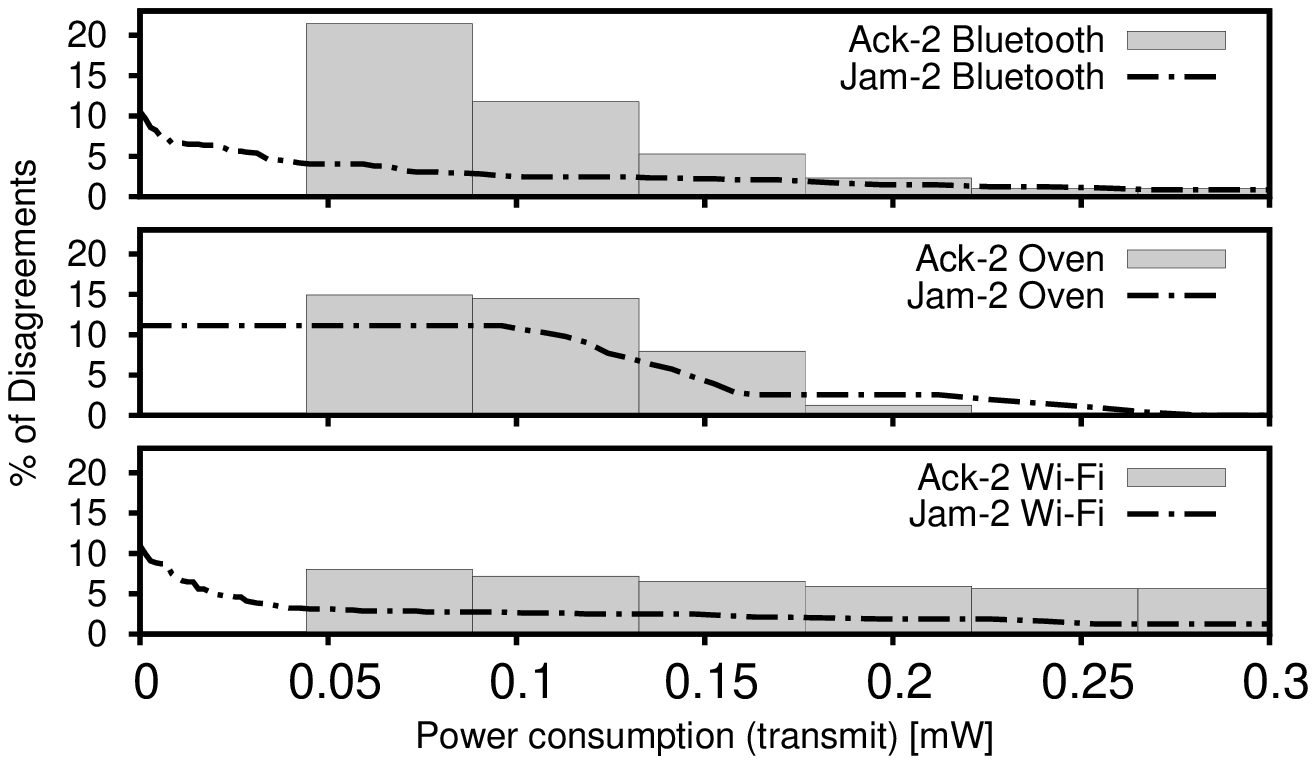}
			\label{fig:JAM2_ACK2_disagreements_energy}
		}
		\vspace{-4.5mm}
		\caption{Compared to Ack-2, Jam-2 achieves less disagreements faster at lower energy cost.}
		\vspace{-5.5mm}
 		\label{fig:ACK2_JAM2_disagreements_2}
	\end{center}
\end{figure*}

Figure~\ref{fig:tx_delay} shows the results. For a payload of one byte (i.e.,
for a generic ACK message), we obtain a transmission delay of 782 $\mu$s and a
processing + sending time of 2083 $\mu$s. Based on these two values, we plot the
fraction of disagreements under different interference patterns as a function of
the time to complete the agreement
(Figure~\ref{fig:JAM2_ACK2_disagreements_time}) and as a function of transmit
power consumption (Figure~\ref{fig:JAM2_ACK2_disagreements_energy}). While we
can continuously vary $t_{jam}$ in Jam-2, for Ack-2 we can only vary
the number $T$ of ACK messages in the train, resulting in discrete boxes in the
figures that are drawn to scale. We see that Jam-2 results in a substantially
smaller fraction of disagreements compared to Ack-2 for a given
time-to-completion and for a given power consumption.

%

%

We also evaluate the performance of Jam-2 under real interference by running an
experiment in our second testbed, deployed in an environment rich of Wi-Fi
traffic and noise generated by microwave ovens. We implement a program that
carries out continuous handshakes both using Ack-2 and using Jam-2 (the latter
uses a $t_{jam} = 2$ ms), and run it over several days, carrying out more than 1
million handshakes.

\begin{figure}[t!]
	\begin{center}	
		\includegraphics[width=0.35\textwidth]{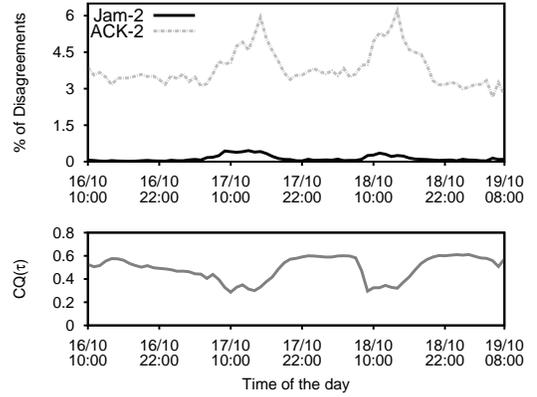}
		\vspace{3.5mm}
		\caption{Long-term experiment comparing the performance of Jam-2 and Ack-2
		under real interference.}
		\vspace{-7mm}
		\label{fig:JAM2_time_2}
	\end{center}
\end{figure}

Figure~\ref{fig:JAM2_time_2} shows the results of our experiment. The top figure
shows the performance of Ack-2 and Jam-2 in terms of the fraction of
disagreements. The bottom figure shows the Channel Quality metric (CQ), that
provides an estimate of interference by capturing a channel's availability over
time~\cite{noda11ipsn_metric}. The figure shows that Jam-2 has a very low
fraction of disagreements compared to Ack-2. Both Jam-2 and Ack-2 have the worst
performance around noon during weekdays (Oct. 16 was a Sunday). During these
times, the fraction of disagreements for Jam-2 grows to 0.5\%, while it is close
to 0\% at other times. We attribute this bad performance during lunch to the
microwave ovens in the nearby kitchen ($t_{jam} < BUSY_{oven}$). The bottom
graph shows that the channel quality is actually lower when the fraction of
disagreements is high, i.e., during noon on weekdays.


%% file: evaluations_unicast_3.tex
\vspace{-1mm}
\subsection{3-way Handshakes: Ack-3 and Jam-3} \label{subsec:evaluation_unicast_jam3}
\vspace{-1mm}
We evaluate the performance of Jam-3 in our JamLab-based testbed and compare it
against a traditional packet-based 3-way handshake Ack-3. The main difference
between Jam-2 and Jam-3 is that in Jam-3 the receiver uses the RSSI of the
initiator's first packet $r_{s}$ to improve the detection of the subsequent
jamming signal as described in Sect.~\ref{subsec:jam_3}. Our focus is on
understanding the impact of the comparison threshold $\Delta_{r}$ on the
probability of positive agreement.

\begin{figure*}[t!]
	\begin{center}
		\subfigure[Ack-3 vs. Jam-3]{
			\includegraphics[width=0.38\textwidth]{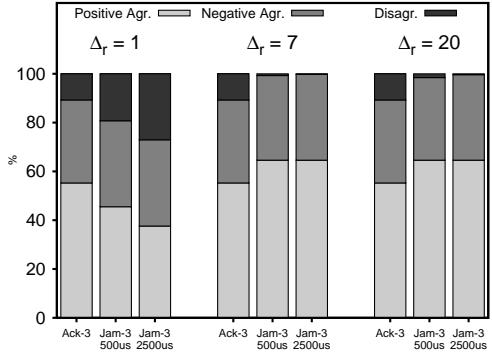}	
			\label{fig:ACK3JAM3}	
		} 		
		\subfigure[Role of $r_{s}$]{
			\includegraphics[width=0.38\textwidth]{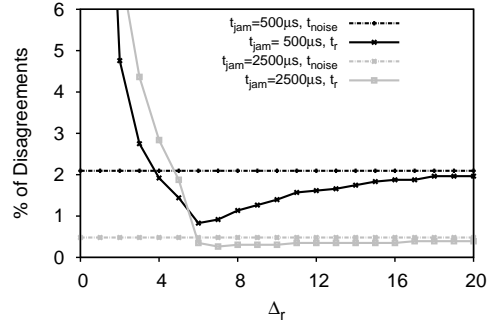}
			\label{fig:JAM3_RSSI}		
		}				
		\vspace{-1.5mm}
		\caption{Performance of Jam-3 and Ack-3 under Bluetooth interference (a), and
		role of $\Delta_{r}$ for Jam-3's efficiency (b).}
		\vspace{-6mm}
 		\label{fig:Ack-3_vs_Jam-3}
	\end{center}
\end{figure*}

Due to space constraints, we only show results for Bluetooth in
Figure~\ref{fig:ACK3JAM3}. The figure shows that Jam-3 is more efficient than
Ack-3 in that Jam-3 has less disagreements and more positive agreements for
$\Delta_{r}=7$dBm  and $\Delta_{r}=20$dBm. This is the case for both jamming
windows $t_{jam}$.  The reason for this superior performance is the robustness
of the jamming sequence detection under interference compared to the robustness
of a normal packet.  For $\Delta_{r}=1$dBm, the figure shows superior
performance for Ack-3. Moreover, it also shows better performance for Jam-3 with
$t_{jam}=500us$ than for Jam-3 $t_{jam}=2500us$. The reason for this lies in the
variation of the received signal strength during the jamming signal. Selecting
small $\Delta_{r}$, increases the probability that a sampled RSSI
value $x_i$ gets below the threshold $(r_{s}-\Delta_{r})$ (see
Section~\ref{subsec:jam_3}). 

Figure~\ref{fig:JAM3_RSSI} depicts the percentage of disagreements as a function
of $\Delta_{r}$. Note that the performance of Jam-3 using $r_{noise}$ as the
comparison threshold (horizontal lines in the figure) is equivalent to that of
Jam-2. For $0 \le \Delta_{r} \le 4$ dBm (for $t_{jam}=500 \mu$s) and $0 \le
\Delta_{r} \le 5$ dBm (for $t_{jam}=2500 \mu$s), Jam-3 performs worse than
Jam-2. This is due to the variations in the RSSI values, that can be up to 5 dBm.

When $\Delta_{r}$ is between 6 and 7 dBm, instead, the number of disagreements
is minimized and reaches between 0 and 1\%. Our logs show that 6 to 7 dBm is a
safety threshold to cover the variations of the signal strength when receiving
the jamming sequence and hence optimizes performance. For larger $\Delta_{r}$
values, the performance is similar to the one obtained using $r_{noise}$ (as in
Jam-2), which slightly increases the percentage of disagreements.

\vspace{1.5mm}

%% file: jamx_broadcast.tex
\section{Jam-B: Broadcast Agreement} \label{sec:jamx_broadcast}

In certain cases, e.g., when agreeing on time slots in a MAC protocol, a
node $S$ needs to agree with a set of $r>1$ neighboring nodes. By exploiting
jamming and following the principles described in
Section~\ref{sec:jamx_unicast}, we design Jam-B, a protocol to reach
agreement in broadcast scenarios.

\begin{figure}[t]
	\begin{center}
		\includegraphics[width=0.44\textwidth]{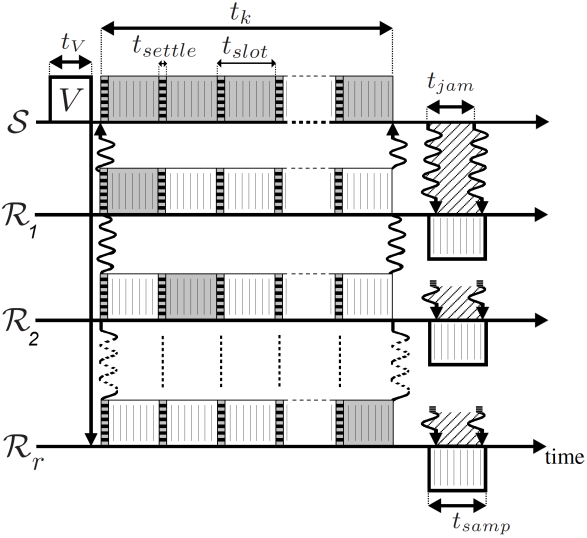}
		\vspace{-5mm}
		\caption{Illustration of Jam-B.}
		\vspace{-6mm}
 		\label{fig:drawing_JAM-B}
	\end{center}
\end{figure}

Jam-B performs a 3-way handshake as depicted in
Figure~\ref{fig:drawing_JAM-B}. First, $\mathcal{S}$ broadcasts
the initial packet containing $V$ to its $r$ neighbors
$\mathcal{R}_{i}$. Then, each of these nodes computes a bit vector
$BV$ holding $k \ge r$ bits as a function of its node address.  This
bit vector is then interpreted as a TDMA schedule with $k$ slots, that
is, a node jams during slot $i$ if and only if $BV[i] = 1$. The
encoding of the bit vectors must be such that the overlay of all nodes
$\mathcal{R}_{i}$ jamming simultaneously results in a continuous
jamming signal at $\mathcal{S}$. As in Jam-2, $\mathcal{S}$ samples
RSSI during all slots. If a jamming signal is detected in all slots,
$\mathcal{S}$ jams in turn for $t_{jam}$ and all neighbors sample RSSI
to detect this jamming signal.

Besides the choice of the duration $t_{slot}$ of one jamming slot, the
encoding of the bit vectors affects the detection accuracy. In the
simplest case, the number of slots equals the number of neighbors and
every neighbor has exactly one bit set to 1 in its bit vector. However,
a better alternative is to use more slots than neighbors (possibly
trading off more slots for a shorter slot duration $t_{slot}$) and use
an encoding that sets multiple bits to 1 in each neighbors' bit
vector maximizing the distance between the bits. This
approach reduces the chances that interference affects all the
jamming slots of a node. We leave the derivation of an optimal
encoding scheme for future work.

\vspace{3mm}

\subsection{Jam-B Implementation} \label{subsec:implementation_JAM-B}

We implement Jam-B on the same platforms described in
Section~\ref{subsec:implementation}. In our implementation, the bit
vector consists of 16 bits. A slot has a fixed duration $t_{slot} = 1000
\mu$s, which corresponds to roughly 49 RSSI samples.  Only a subset
of those samples is used by $\mathcal{S}$ to assess the presence of
jamming due to synchronization inaccuracies. While the initial
broadcast message acts a synchronization beacon for the nodes, packet
processing delays, radio settling times, and RSSI readout latency
(about 21 $\mu$s) lead to variable time offsets among the nodes. Also,
slot sizes are not an integral multiple of the RSSI sampling
interval. Finally, the RSSI values need a time of about 128 $\mu$s to
settle. Therefore, we introduce a guard time $t_{settle}$ of 8 RSSI
samples or 169 $\mu$s between slots during which RSSI samples are
discarded. We use $t_{jam} = 2$ ms.

For the bit vector encoding, we chose a scheme where each node $\mathcal{R}_{i}$
sets $\lfloor 16/r \rfloor$ bits to 1 in its bit vector such that no pair of
nodes has a one bit at the same position. Each of the
remaining $16 - \lfloor 16/r \rfloor$ bits is allocated to multiple nodes in a
symmetric fashion. The positions of the bits to 1 in each $BV[i]$ are chosen
such that their distance is maximized.

\vspace{+1.0mm}

\subsection{Evaluation} \label{subsec:evaluation_JAM-B}

\begin{figure}[t!]
	\begin{center}	
		\includegraphics[width=0.38\textwidth]{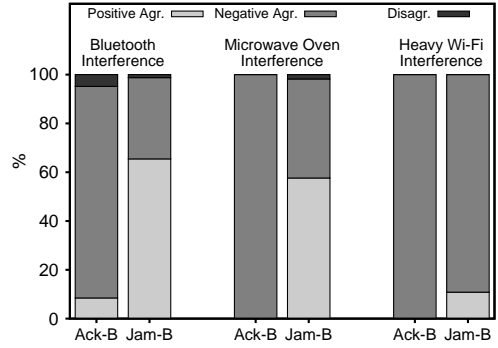}
		\vspace{-2mm}
		\caption{Jam-B significantly outperforms Ack-B, maximizing the amount of
		positive agreements.}
		\vspace{-4mm}
 		\label{fig:ACKB_JAMB_interference}		
	\end{center}
\end{figure}

We evaluate Jam-B using the same experimental setup described in
Section~\ref{subsec:experimental_setup}. We select 7 motes from the JamLab
testbed, one of them as the sender $S$, and the remaining 6 as receivers. We run
the experiment under the different interference patterns used previously. We
compare Jam-B with its equivalent Ack-B implemented using ACK packets instead of
jamming. Figure~\ref{fig:ACKB_JAMB_interference} shows the results, where Jam-B
clearly outperforms Ack-B. The latter leads to a very high number of negative
agreements because of the high probability that at least one of the packets
involved in the exchange gets corrupted. The gain obtained using Jam-B is
therefore much bigger compared to simple unicast scenarios.


%% file: related.tex
\vspace{-1.5mm}
\section{Related Work} \label{sec:related}
\vspace{-0.5mm}

Agreement is a well-known problem in distributed systems. Pioneering work in the
late 1970s highlighted the design challenges when attempting to coordinate an
action by communicating over a faulty channel~\cite{akkoyunlu75constraints,
gray78database}.

In the context of wireless sensor networks, the agreement problem has
not been been widely addressed. The main focus has been on security
for the exchange of cryptographic keys~\cite{du2004Infocom}, and on
average consensus for nodes to agree on a common global value after
some iterations~\cite{xiao2005IPSN}. Similarly to these studies, our
work aims at protocols that allow a set of nodes to agree on a piece
of information, but our study has two distinctive characteristics, we
focus: (i) on overcoming the effects of common interference signals
and (ii) on solutions that fit applications with timeliness requirements.
 
Our work is motivated by studies reporting the degrading quality of service
caused by the overcrowding of the RF spectrum, in particular in unlicensed bands
such as the 2.4 GHz band~\cite{zhou06crowded}. Several studies have proposed
solutions to overcome these interference problems. Chowdhury and Akyildiz
identify the type of interferer in order to adapt the resources accordingly and
reduce packet losses~\cite{chowdhury09classification}. 
Liang et al. increase the resilience of packets challenged by Wi-Fi interference
using multi-headers and FEC techniques~\cite{liang10surviving}. Other studies
have proposed to hop out of interfered channels dynamically. The
ARCH~\cite{sha11rtas} and Chrysso~\cite{iyer2011chrysso} protocols switch the
communication frequency when interference is detected. These protocols require
algorithms for measuring and quantifying interference and there are several
candidates in the literature~\cite{musaloiu08minimising, noda11ipsn_metric,
xing09interference}. On the same line of thought, cognitive radio techniques,
which aim at exploiting unused RF spectrum, have been explored in sensor
networks~\cite{ansari10agile}. All these studies rely on
packet exchanges to coordinate the channel switching and Jam-X can be utilized
to improve their performance.

Another set of studies propose to cope with interference by exploiting
its idle times. Hauer et al. report the interference observed by a
mobile body area network in public spaces, and the study shows the
intermittent interference caused by Wi-Fi spots in all IEEE~802.15.4
channels~\cite{hauer09interference}.  Similarly, Huang et al. have
shown that Wi-Fi traffic inherently leaves ``a significant amount of
white spaces'' between 802.11 frames~\cite{huang10white}. Similarly to
these studies, Jam-X exploits idle times for data packets, but Jam-X
also takes advantage of the bursty non-continuous nature of
interference to identify jamming signals that are part of a handshake.

In terms of agreements for broadcast scenarios, there is an interesting set of
related studies. SMACK~\cite{dutta09smack} uses simultaneous transmissions from
a number of nodes to implement reliable acknowledgements. The nodes use
orthogonal frequency division modulation radios (OFDM) to receive packets on
different subcarriers simultaneously. These sophisticated radios are able to
identify the sender of the message, which is not possible with the
single-channel radios implementing 802.15.4. In contrast to SMACK,
Pollcast~\cite{demirbas08singlehop}, Countcast~\cite{demirbas08singlehop} and
Backcast~\cite{dutta08wireless} approximate the number of senders but they cannot
identify individual addresses, which is necessary for an agreement.
A study similar in spirit to Jam-X is provided by Krohn et al., where the authors
exploit jamming signals for estimating the number of one-hop
neighbors~\cite{krohn04sdjs}. Similar to Countcast, their scheme results in
an approximation, which is insufficient to reach agreement.



%% file: conclusions.tex
\section{Conclusions} \label{sec:conclusions}

In this study we propose Jam-X, a simple and efficient agreement
protocol for wireless sensor networks working under
interference. Jam-X introduces a novel technique that utilizes jamming
signals as binary acknowledgments of the reception of a packet. The
key insight of our work is a mechanism that permits the reliable
detection of \textit{intended} jamming signals in the presence of different
interference patterns. This method overcomes the fundamental
limitations of regular acknowledgment packets which often cannot be
successfully decoded when interference is present. 


We implement Jam-X using Contiki on 802.15.4
platforms and evaluate it with different sources of interference.
Our results show that Jam-X outperforms traditional methods using packet-based
acknowledgements for both unicast and broadcast scenarios. For a 2-way unicast
handshake, Jam-2 sustains 90\% and 35\% positive agreements under microwave
ovens interference and heavy Wi-Fi traffic, respectively. A traditional 2-way
handshake (Ack-2), instead, sustains only 60\% and 10\%, respectively, requiring
longer delays and higher power consumption.

\vspace{-1mm}
%
